\documentclass[11pt,a4paper]{article}
\usepackage{jcappub}
\usepackage{aas_macros}
\usepackage{color}
\usepackage{tocloft}
\newlistof[section]{footnotes}{fnt}{\listfootnotesname}

\let\oldfootnote\footnote 
\renewcommand\footnote[1]{%
    \refstepcounter{footnotes}%
    \oldfootnote{#1}%
    \addcontentsline{fnt}{footnotes}{\protect
\numberline{\thefootnotes}#1}%
}
\usepackage{graphicx} 
\title{Formulation and constraints on decaying dark matter with finite mass daughter particles}

\author[1]{Shohei Aoyama}
\author[1]{Kiyotomo Ichiki}
\author[1]{Daisuke Nitta}
\author[1,2,3]{and Naoshi Sugiyama}
\affiliation[1]{Department of Physics and Astrophysics, Nagoya University,\\
Furo-cho,Chikusa ward,Nagoya 464-8602, Japan}
\affiliation[2]{Institute for the Physics and Mathematics of the Universe (IPMU),\\
The University of Tokyo, Kashiwa, Chiba, 277-8568, Japan}
\affiliation[3]{Kobayashi-Maskawa Institute for the Origin of Particles and the Universe,\\
Nagoya  University,Furo-cho, Chikusa ward, Nagoya 464-8602, Japan}

\emailAdd{aoyama314@a.phys.nagoya-u.ac.jp}

\abstract{ Decaying dark matter cosmological models have been proposed
to remedy the overproduction problem at small scales in the standard
cold dark matter paradigm.  We consider a decaying dark matter model in
which one CDM mother particle decays into two daughter particles, with
arbitrary masses. A complete set of Boltzmann equations of dark matter
particles is derived which is necessary to calculate the evolutions of
their energy densities and their density perturbations.  By comparing
the expansion history of the universe in this model and the
free-streaming scale of daughter particles with astronomical
observational data, we give constraints on the lifetime of the mother
particle, $\Gamma^{-1}$, and the mass ratio between the daughter and the
mother particles $m_{\rm D}/m_{\rm M}$.  From the distance to the last
scattering surface of the cosmic microwave background, we obtain
$\Gamma^{-1}>$ 30 Gyr in the massless limit of daughter particles and,
on the other hand, we obtain $m_{\rm D} >$ 0.97$m_{\rm M}$ in the limit
$\Gamma^{-1}\to 0$. The free-streaming constraint tightens the bound on
the mass ratio as $\left(\Gamma^{-1}/10^{-2}{\rm Gyr}\right) \lesssim
\left((1-m_{\rm D1}/m_{\rm M})/10^{-2}\right)^{-3/2}$ for $\Gamma^{-1} <
H^{-1}(z=3)$.}  
\arxivnumber{1106.1984}

\begin{document}
\maketitle

\section{Introduction}
There are numerous astronomical phenomena which indicate the existence
of cold dark matter (CDM), such as flat rotation curves of galaxies,
anisotropies of cosmic microwave background (CMB), and measurements of
gravitational lenses \cite{Cold..Dark..Matter}. However, we know little
about the nature of cold dark matter. In fact, at sub-galactic scales, 
there are some discrepancies between CDM model predictions by numerical
simulations and astronomical observations, such as the abundance of
substructures in dark matter halos and the inner matter distribution in
galaxies \cite{1999ApJ...524L..19M,2009ApJ...696.2115I}.

The discrepancies have stimulated numerous proposals to rescue the
standard CDM model both from astrophysics and particle physics
communities.  One of the astronomical explanations for these
discrepancies is gas feedback: the heated gas by astrophysical processes
such as supernova explosions may alter the matter distribution at the
inner core of galaxies and/or hinder the galaxy formation in
sub-halos. Explanation from particle physics, on the other hand,
includes decaying dark matter, interacting dark matter, warm dark matter
and so on (for a review, see \cite{2003Sci...300.1909O}).  Among them,
Cen \cite{2001ApJ...546L..77C} proposed the decaying dark matter model
to solve both the overproduction problem of dwarf galaxies and the
over-concentration problem of the inner core.

There are many kinds of studies about decaying dark matter, in
particular, constraining its lifetime $\Gamma^{-1}$
\cite{2004PhRvL..93g1302I,2010PhRvD..81h3511P,2003ApJ...597..645O,2010PhRvD..81j3501P,2005PhRvD..72f3510K,1999PhRvD..60l3508K,2001ApJ...546L..77C,1993NuPhB.403..671K,2009JCAP...06..005D,2008PhRvD..77b3009A,2005PhR...405..279B,2005PhLB..625....7K,2008PhLB..665...50P}. For
example, the light element abundances predicted by Big-Bang
Nucleosynthesis set the limit on the lifetime if the lifetime is relatively short
\cite{1993NuPhB.403..671K,2005PhLB..625....7K}.  For decaying dark
matter with much longer lifetime, constraints can be set from the other
cosmological observations, such as anisotropies of CMB, the abundance of
clusters of galaxies, halo mass profiles and so on
\cite{2009JCAP...06..005D,2008PhRvD..77b3009A,2003ApJ...597..645O,2005PhR...405..279B,2004PhRvL..93g1302I}.
Most of these constraints have been obtained with an assumption that
dark matter decays into massless daughter particles.  
Recently, however,  a decaying dark matter model is proposed in which the mass
difference between the mother and daughter particles is very small and
the lifetime of the mother particle is as long as the age of the
universe \cite{2011PhRvD..83f3504B,2010PhRvD..81h3511P}.
By using
N-body simulations of galaxy halos, Peter \cite{2010PhRvD..81h3511P} put constraints on $\Gamma^{-1}$ and the mass
of daughter particles through the recoil speed of daughter particles
$v_{\rm k}$ in the case where the mass difference of the mother and one
of the daughter particles is small. There are several theoretical
proposals for models with decaying dark matter (see, eg.,
\cite{1998PhRvL..81.4048C,1999PhRvD..59d7301B,2001PhRvL..86..954L,2002PhLB..527...18K,2008PhLB..665...50P}).

In this paper we extend previous works
\cite{2004PhRvL..93g1302I,2009JCAP...06..005D} by taking into account
the finite mass effects of daughter particles with arbitrary masses. We
derive a set of Boltzmann equations to describe this decaying dark
matter scenario, and study the effects of the decay on cosmology.  In
our formulation, the masses of the mother and two daughter particles,
$m_{\rm M},m_{\rm D1},$ and $m_{\rm D2}$, can be set to arbitrary
values. {For an example we assume that the mother particle is Next Lightest Supersymmetric Particle (NLSP), which belongs to Minimal Supersymmetric
Standard Model(MSSM). If mother is a NLSP, its mass should be larger
than at least 46 GeV \cite{Nakamura:2010zzi}
\footnote{{If the condition is relaxed, it is  possible
that the mass of mother particle is lighter than the mass of the
lightest particle which belongs to MSSM (see eg.,\cite{2003PhRvD..67f3519B,2003PhRvD..68d3506B,2010arXiv1009.0549B}). However,
we can still regard the mother particle massive enough in these cases to
take the one of dauther particles as massless one.}}.} Due to the R-parity
conservation, if the mother particle is a SUSY particle, one of the
daughter particles is also a SUSY particle and the other is a standard
model particle. We only consider the possibility that the created
standard model particle is a lepton, which can be considered effectively
massless in the decay process. In this paper, therefore,
{when we give constraints on the lifetime of the mother
particle we assume that one of the daughter particles is massless.}  By
 solving
the set of Boltzmann equations for the mother and daughter particles and
by integrating the distribution functions we can derive the time
evolution of the energy density and the typical free-streaming scale of
daughter particles. We then give constraints on $\Gamma ^{-1}$ and the
mass ratio $m_{\rm D1}/m_{\rm M}$ from the "comoving" angular diameter distances to
CMB and the position of Baryon Acoustic Oscillation (BAO).  We also give
constraints from the free-streaming scale of daughter particles by
comparing with the Lyman $\alpha $ data
\cite{2003AJ....125...32B,2009JCAP...05..012B}.

This paper is organized as follows. In Sec. 2, we derive a complete set of
Boltzmann equations for mother and daughter particles to describe the
evolutions of their energy densities and density perturbations including
the decay process. In Sec. 3 we solve the background distribution
functions of mother and daughter particles. 
We then show the evolution of their energy densities and give observational
constraints from astronomical data sets on the lifetime of the mother
particle and the mass ratio in Sec. 4. Finally Sec. 5 is devoted to our conclusion.
Throughout this paper the speed of light $c$ is set to unity.

\section{Formulation}

 {In this section, we aim to derive a set of equations
for mother and daughter particles in order to describe the time evolutions of
their energy densities and their associated linear density perturbations
including the decay process. In order to do this, we write down a
complete set of Boltzmann equations (both at zero-th order and first order) in
this section and in Appendix C.  The background equations 
in Sec. 2 will be indeed solved in our analysis to derive
constraints from the geometry of the universe, i.e. from BAO and CMB
peak positions, and Hubble constant today.}
 
To describe the decay process, we define the decay rate $\Gamma(q_{\rm
D},q_{\rm M})$, which is the function describing how many daughter
particles with comoving momentum $q_{\rm D}$ are created for a unit time
interval from the mother particles with momentum $q_{\rm D}$. First, let
us write the Boltzmann equations for the distribution functions of
mother and daughter particles. The Boltzmann equation for the
distribution function of the mother particles $f_{\rm M}(q_{\rm M})$ is
\begin{equation} 
\dfrac{df_{\rm M}}{dt}=\dfrac{\partial f_{\rm M}}{\partial t  }+\dfrac{d x^{i}}{dt}\dfrac{\partial f_{\rm M}}{\partial x^{i}}+\dfrac{d q_{\rm M}}{dt}\dfrac{\partial f_{\rm M}}{\partial q_{\rm M}}+\dfrac{d n_{i}}{dt}\dfrac{\partial f_{\rm M}}{\partial n_{i}}=\bigg(\dfrac{\partial f_{\rm M}}{\partial t}\bigg)_{C}~,\label{eq:2.1}
\end{equation}
where $n_{i}$ is the unit vector in the direction of the momentum.

The collision term in eq.(\ref{eq:2.1}) can be expressed as an
integration of $\Gamma (q_{\rm D},q_{\rm M})f_{\rm M}$ with $q_{\rm D}$
as
\begin{equation} 
\bigg(\dfrac{\partial f_{\rm M}}{\partial t}\bigg)_{C}=-\displaystyle\int \Gamma (q_{\rm D},q_{\rm M}) f_{\rm M} d^{3}q_{\rm D}~.
\end{equation}
Similarly, the Boltzmann equations of daughter particles are
\begin{equation} 
\dfrac{df_{{\rm D}j}}{dt}=\dfrac{\partial f_{{\rm D}j}}{\partial t}+\dfrac{d x^{i}}{dt}\dfrac{\partial f_{{\rm D}j}}{\partial x^{i}}+\dfrac{d q_{\rm D}}{dt }\dfrac{\partial f_{{\rm D}j}}{\partial q_{\rm D}}+\dfrac{d n_{i}}{dt}\dfrac{\partial f_{{\rm D}j}}{\partial n_{i}}=\bigg(\dfrac{\partial f_{{\rm D}j}}{\partial t  }\bigg)_{C}~,\label{eq:2.3}
\end{equation}
where $j$ is the particle index $j=1,2$.

The daughter particles in this model are created only through the decay
of mother particles. Thus the collision term can be written as
\begin{equation} 
\bigg(\dfrac{\partial f_{{\rm D} j}}{\partial t}\bigg)_{C}=+\displaystyle\int \Gamma (q_{\rm D},q_{\rm M}) f_{\rm M} d^{3}q_{\rm M}~.
\end{equation}

Secondly, we derive $\Gamma (q_{\rm D},q_{\rm M})$ explicitly. From the
definition of CDM, the thermal motion of mother particles is so slow
that we can regard them as objects at rest. To be specific, the kinetic
energy of mother particles is much smaller than the mass deficit of
the decay. Due to the conservation of the momentum, two daughter
particles are emitted in the opposite directions and the amplitudes of
these momenta are the same. So we can write the decay rate $\Gamma
(q_{\rm D},q_{\rm M})$ as a function proportional to the delta function
of $q_{\rm D}$, which is determined by the energy momentum
conservation. 
From the conservation, $q_{\rm M}$ and $q_{\rm D}$ should satisfy
the condition as
\begin{equation} 
\sqrt{m_{\rm M}^{2}a^{2}+q_{\rm M}^{2}}=\sqrt{m_{\rm D1}^{2}a^{2}+q_{\rm D}^{2}}+\sqrt{m_{\rm D2}^{2}a^{2}+q_{\rm D}^{2}}~.\label{eq:2.8}
\end{equation}
Because $q_{\rm M}\ll am_{\rm M}$, the solution of eq.(\ref{eq:2.8})
leads to 
\begin{equation} 
q_{\rm D}\simeq \dfrac{1}{2}\sqrt{A_{\rm M}a^{2}+B_{\rm M} q_{\rm M}^{2}}~.
\end{equation}
In addition, we assume that these daughter particles are
isotropically emitted. By taking into account the spherical symmetry of
the decay, the decay rate should be written as,
\begin{equation} 
\Gamma (q_{\rm D},q_{\rm M})=\dfrac{\Gamma }{4\pi q_{\rm D}^{2}}\delta \left( q_{\rm D} - \dfrac{1}{2}\sqrt{A_{\rm M}a^{2}+B_{\rm M} q_{\rm M}^{2}} \right)~,
\end{equation}
where $\Gamma^{-1}$ is the lifetime of the mother particle and $a$ is
the scale factor. Here $A_{\rm M}$ and $B_{\rm M}$ are constants defined
as
\begin{eqnarray} 
A_{\rm M} &\equiv & m_{\rm M}^{2}-2(m_{\rm D1}^{2}+m_{\rm D2}^{2})+\dfrac{(m_{\rm D1}^{2}-m_{\rm D2}^{2})^{2}}{m_{\rm M}^{2}}~,\\
B_{\rm M} &\equiv & 1- \dfrac{(m_{\rm D1}^{2}-m_{\rm D2}^{2})^{2}}{m_{\rm M}^{4}}~.
\end{eqnarray}

Thirdly, we consider the first order perturbations of the distribution
functions $f_{\rm M}$, $f_{{\rm D}j}$ $(j=1,2)$ as follows. It is
convenient to write a distribution function as a zero-th order
distribution, which is the background distribution, plus a
perturbed function $\Psi$,
\begin{eqnarray} 
f_{\rm M}&\equiv &f_{\rm M}^{(0)}(q_{\rm M},t)(1+\Psi_{\rm M}(x^{i},q_{\rm M},n_{i},t)),\label{eq:2.10}\\
f_{{\rm D}j}&\equiv &f_{{\rm D}j}^{(0)}(q_{\rm D},t)(1+\Psi_{{\rm D}j}(x^{i},q_{\rm D},n_{i},t)). \label{eq:2.11}
\end{eqnarray}
By substituting eq.(\ref{eq:2.10}) and (\ref{eq:2.11}) into
eq.(\ref{eq:2.1}) and (\ref{eq:2.3}), respectively, and comparing the
equations order by order, we obtain the following equations at zero-th order
for the mother particles as
\begin{eqnarray} 
\text{unperturbed} &:&\dot{f}_{\rm M}^{(0)}=-\displaystyle\int \Gamma (q_{\rm D},q_{\rm M}) f_{\rm
 M}^{(0)} d^{3}q_{\rm D}=- \Gamma f_{\rm M}^{(0)}~.\label{2.12}
\end{eqnarray}
Here overdot denotes the derivative with respect to the cosmic time $t$.

For daughter particles, we obtain the following equation,
\begin{eqnarray} 
\text{unperturbed} &:&\dot{f}_{{\rm D}j}^{(0)}=\displaystyle\int \Gamma
 (q_{\rm D},q_{\rm M}) f_{\rm M}^{(0)}(q_{\rm M}) d^{3}q_{\rm M}=\dfrac{4\Gamma }{B_{\rm M}}\dfrac{q_{\rm D}^{\prime }}{q_{\rm D}}f^{(0)}_{\rm M}(q^{\prime }_{\rm D}) ~, \label{2.14}
\end{eqnarray}
where 
\begin{equation} 
q_{\rm D}^{\prime }\equiv \sqrt{\dfrac{4 q_{\rm D}^{2}-A_{\rm M}a^{2}}{B_{\rm M}}}~.
\end{equation}
The first order equations are presented in Appendix C.

{The two unperturbed equations above can be interpreted as follows. The
equation (\ref{2.12}) states that all the mother particles should decay
with the decay rate $\Gamma$. For daugher particles, on the other hand,
the equation (\ref{2.14}) means that the daugher particles with momentum
$q_{\rm D}$ should be created at a rate proportional to $\Gamma f_{\rm
M}^{(0)}(q_{\rm D}')$, where $q_{\rm D}'$ is the momentum which the
mother particles should have for the created daugher particle to have
the momentum $q_{\rm D}$ after the decay.} Note that because from
eq. (\ref{2.14}) the unperturbed distribution functions of two dauther
particles coincide, $f^{(0)}_{{\rm D}1}=f^{(0)}_{{\rm D}2}$, and
hereafter we simply denote them as $f^{(0)}_{\rm D}$.

\section{Calculation of background distribution functions}
\subsection{Mother particle} 
The temperature of CDM is low compared with the rest energy of the
particle. If the mother particles do not decay, i.e.  $\Gamma =0 $, the
background distribution function of the particles is given by the
Maxwell-Boltzmann function. In this case, or in the very early universe
where the decay is negligible, the background distribution $\tilde{f}_{\rm
M}^{(0)}$ is
\begin{equation} 
\tilde{f}_{\rm M}^{(0)}(q_{\rm M},t)=\dfrac{1}{(2\pi m_{\rm M}T_{\rm M0})^{3/2}}\exp \left( -\dfrac{q_{\rm M}^{2}}{2 m_{\rm M}T_{\rm M0}} \right)~,
\end{equation}
where $T_{\rm M0}$ is the present temperature of mother particles. We consider $m_{\rm M}=1.0$ TeV as a working example. Because of the calculation performed in Appendix b, we obtain the temperature of mother particles as
\begin{equation} 
T_{\rm M0} \simeq 1.7\times 10^{-14}~\text{K}=1.4\times 10^{-18}~\text{eV}{}~,
\end{equation}

And we set the Boltzmann constant $k_{\rm B} =1$. In fact, the result depends only on
the mass ratio $m_{\rm D1}/m_{\rm M}$, but not on the absolute value of
$m_{\rm M}$.  Including the decay, $f_{\rm M}^{(0)}$ is given by the
solution of eq.(\ref{2.12}) as
\begin{equation} 
f_{\rm M}^{(0)}(q_{\rm M},t)=\dfrac{1}{(2\pi m_{\rm M}T_{\rm M0})^{3/2}}\exp \left( -\dfrac{q_{\rm M}^{2}}{2 m_{\rm M}T_{\rm M0}} -\Gamma t \right)~. \label{eq.3.2}
\end{equation}
Here we have normalized the distribution function at $t=0$ as 
\begin{equation} 
\displaystyle\int_{0}^{+\infty }4 \pi q_{\rm M}^{2} f_{\rm M}^{(0)}(q_{\rm M},t=0)d q_{\rm M}=1~.
\end{equation}
The number density of the mother particles, $n_{\rm M}$, is determined by
\begin{equation} 
n_{\rm M}=\rho_{\rm c}\Omega_{\rm DM}/m_{\rm M}~,
\end{equation} 
where $\rho_{\rm c}$ is the critical density of the universe,
$\Omega_{\rm DM} $ is the density parameter of dark matter normalized by
the critical density at present. Note that the parameter $\Omega_{\rm
DM}$ in the above equation is an extrapolated value of the density
parameter of mother particles without decay. Then the energy density of
the mother particles $\rho_{\rm M}$ is given by
\begin{equation} 
\rho_{\rm M}=\dfrac{\rho_{\rm c}\Omega_{\rm DM}}{m_{\rm M}a^{4}}\displaystyle\int_{0}^{+\infty }4 \pi q_{\rm M}^{2}\sqrt{m_{\rm M}^{2}a^{2}+q_{\rm M}^{2}}f_{\rm M}^{(0)}(q_{\rm M},t) dq_{\rm M} ~.
\end{equation}

\subsection{Daughter particle}
In this section we derive the background distribution function of
daughter particles. By substituting $f_{\rm M}^{(0)}$ in
eq.(\ref{eq.3.2}) into eq.(\ref{2.14}), we obtain a partial differential
equation for the unperturbed distribution function $f_{\rm D}^{(0)}$ as,
\begin{equation} 
\dot{f}^{(0)}_{\rm D}=\dfrac{\sqrt{2}\Gamma }{\pi^{3/2} q_{\rm D}B_{\rm M}^{3/2}}\Bigg( \dfrac{1}{m_{\rm M}T_{\rm M0}} \Bigg)^{3/2} \sqrt{4q_{\rm D}^{2}-a^{2}A_{\rm M}}\exp \Bigg( -\dfrac{4q_{\rm D}^{2}-a^{2}A_{\rm M}}{2B_{\rm M} m_{\rm M} T_{\rm M0}}\Bigg)\exp\left( -\Gamma t\right)~. \label{3.9}
\end{equation}
Since the daughter particles did not exist in the early stage of the universe,
\begin{equation*} 
f_{\rm D}^{(0)}(q_{\rm D},t=0)=0~.
\end{equation*}
When a mother particle decays, the amplitudes of physical momenta of two
daughter particles are the same, and we denote it as $p_{\rm th}$.
Because the physical momentum of daughter particles decays as $\propto
a^{-1}$ as the universe expands, the daughter particles which are
created in the past should have the comoving momentum smaller than
$p_{\rm th}$. {In addition, since the thermal motions of
mother particles are very slow compared with their mass, the time can be
decided uniquely when
the daughter particles with the momentum $q_{\rm D}$ were
created.  To put it concretely, the redshift
$z_{\rm D}$ which corresponds the redshift when the daughter particles
with the present momentum $q_{\rm D}$ were created should satisfy
\begin{equation*} 
q_{\rm D}=\dfrac{1}{1+z_{\rm D}}p_{\rm th}\sim \dfrac{1}{1+z_{\rm D}}\Delta m~.
\end{equation*}
where $\Delta m$ is the mass difference between the
mother and daughter particles. In the limit {$m_{\rm D1}+m_{\rm D2}\to 0$}, $p_{\rm th}$ is equal to $\Delta m$.}

Now let us consider the time evolution of $f_{\rm D}^{(0)}$ at a fixed
comoving momentum $q_{\rm D}$.  The source term of eq.(\ref{3.9}) is
exponentially suppressed in the very early universe when $a^2 \ll
4q_{\rm D}^2/A_{\rm M}$, because the typical momentum of the daughter
particles $q_{\rm D}$ is much larger than the termperature of the mother
particles $T_{\rm M0}$.  The source term, on the other hand, should be
zero when $a^2 \geq 4q_{\rm D}^2/A_{\rm M}$, which comes from the
energy-momentum conservation law. Therefore, the source is important
only around $t\lesssim t^{*}_{q_{\rm D}}$ where $t^{*}_{q_{\rm D}}$ is
defined by
\begin{equation*} 
4q_{\rm D}^{2}-a^{2}(t_{q_{\rm D}}^{*})A_{\rm M}=0~.
\end{equation*}

\begin{figure}
 \begin{center}
  \includegraphics[width=100mm]{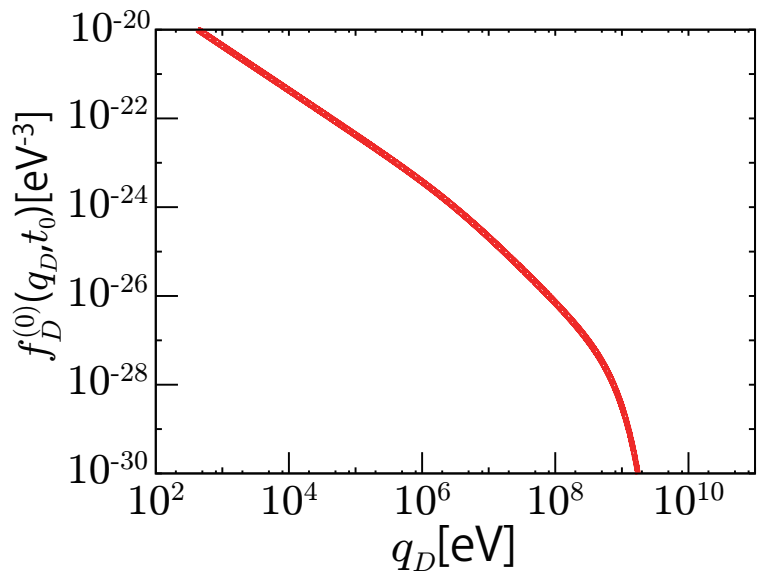}
 \end{center}
  \caption{Distribution function of daughter particles at present
 as a function of comoving momentum of daughter particles
 $q_{\rm D}$, with parameters $m_{\rm M}=1.0$ TeV, $m_{\rm D1} = 0.98$
 TeV, and $\Gamma^{-1}=0.1$ Gyr.}  \label{fig.1}
  \end{figure}

To take advantage of the rapid convergence of the source term of $f_{\rm
D}^{(0)}(q_{\rm D},t)$ we expand $a(t)$ 
around $t_{q_{\rm D}}^{*}$ as
 \begin{equation}
a(t)\simeq a(t^{*}_{q_{\rm D}})+\dot{a}(t^{*}_{q_{\rm D}})(t-t^{*}_{q_{\rm D}})\equiv
 a(t^{*}_{q_{\rm D}})+\dot{a}(t^{*}_{q_{\rm D}})\varepsilon ~~~(\varepsilon \leq 0)~,\\
 \end{equation}
where $\varepsilon = t-t^*_{q_{\rm D}}$.
Here we have omitted the higher order terms in the expansion because the
source of $f_{\rm D}^{(0)}(q_{\rm D},t)$ decays exponentially backward
in time for $t<t_{\rm q_{\rm D}}^{*}$. 
Then the evolution equation (eq.(\ref{3.9})) can be expanded as
\begin{eqnarray}
\dot{f}_{\rm D}^{(0)}(q_{\rm D},t)&=&\dot{f}_{\rm D}^{(0)}(q_{\rm
 D},t_{q_{\rm D}}^{*}+\varepsilon )\simeq \dfrac{\sqrt{2}\Gamma
 }{\pi^{3/2} q_{\rm D}B_{\rm M}^{3/2}}\Bigg( \dfrac{1}{m_{\rm M}T_{\rm
 M0}} \Bigg)^{3/2} \sqrt{-a(t^{*}_{q_{\rm D}})\dot{a}(t^{*}_{q_{\rm
 D}})\varepsilon A_{\rm M}}~, \nonumber \\
& &\times \exp \Bigg( \bigg(\dfrac{A_{\rm M}a(t^{*}_{q_{\rm D}})\dot{a}(t^{*}_{q_{\rm D}})}{B_{\rm M} m_{\rm M} T_{\rm M0}}- \Gamma \bigg) \varepsilon \Bigg)\exp (-\Gamma t^{*}_{q_{\rm D}})~,\label{eq:3.8}
\end{eqnarray}
and the integration of time $t$ can be replaced with that of $\varepsilon $.
Furthermore, we can extend the range of integration as
\begin{equation} 
f_{\rm D}^{(0)}(q_{\rm D},t_{q_{\rm D}}^{*})=\displaystyle\int_{-t^{*}_{q_{\rm D}}}^{0}\dot{f}_{\rm D}^{(0)}(q_{\rm D},t_{q_{\rm D}}^{*}+\varepsilon )d\varepsilon \simeq \displaystyle\int_{-\infty }^{0}\dot{f}_{\rm D}^{(0)}(q_{\rm D},t_{q_{\rm D}}^{*}+\varepsilon )d\varepsilon~. \label{3.11}
\end{equation}
This is because the term in the exponential in eq.(\ref{3.11}) is very large in negative value; $\dfrac{A_{\rm M}a(t^{*}_{q_{\rm
D}})\dot{a}(t^{*}_{q_{\rm D}})}{B_{\rm M} m_{\rm M} T_{\rm M0}}(-t^{*}_{q_{\rm D}})
\sim -{\cal O}(10^{23})\ll
-1$ for $t^{*}_{q_{\rm D}}$ around recombination,  and hence the integration of eq.(\ref{eq:3.8}) for $t=[-\infty,0]$, i.e.,
$\varepsilon=[-\infty,-t_{\rm q_{\rm D}}^{*}]$ is negligible.  
We are
then able to perform this integration of $\varepsilon $ analytically to
obtain
\begin{equation} 
f_{\rm D}^{(0)}(q_{\rm D},t)=\Gamma \dfrac{\sqrt{\pi A_{\rm M} a(t^{*}_{q_{\rm D}})\dot{a}(t^{*}_{q_{\rm D}})} }{B_{\rm M}^{3/2}q_{\rm D}}\left( \dfrac{A_{\rm M}a(t^{*}_{q_{\rm D}})\dot{a}(t^{*}_{q_{\rm D}})}{B_{\rm M}m_{\rm M}T_{\rm M0}} -\Gamma \right)^{-3/2} \Bigg( \dfrac{1}{m_{\rm M}T_{\rm
 M0}} \Bigg)^{3/2}\exp (-\Gamma t^{*}_{q_{\rm D}} )\theta (t-t^{*}_{q_{\rm D}})~.  
\end{equation}
We show the shape of the distribution function of daughter particles
$f_{\rm D}^{(0)}(q_{\rm D},t)$ at present time in figure \ref{fig.1}.

 \begin{figure}[htbp]
   \begin{center}
\includegraphics[width=70mm]{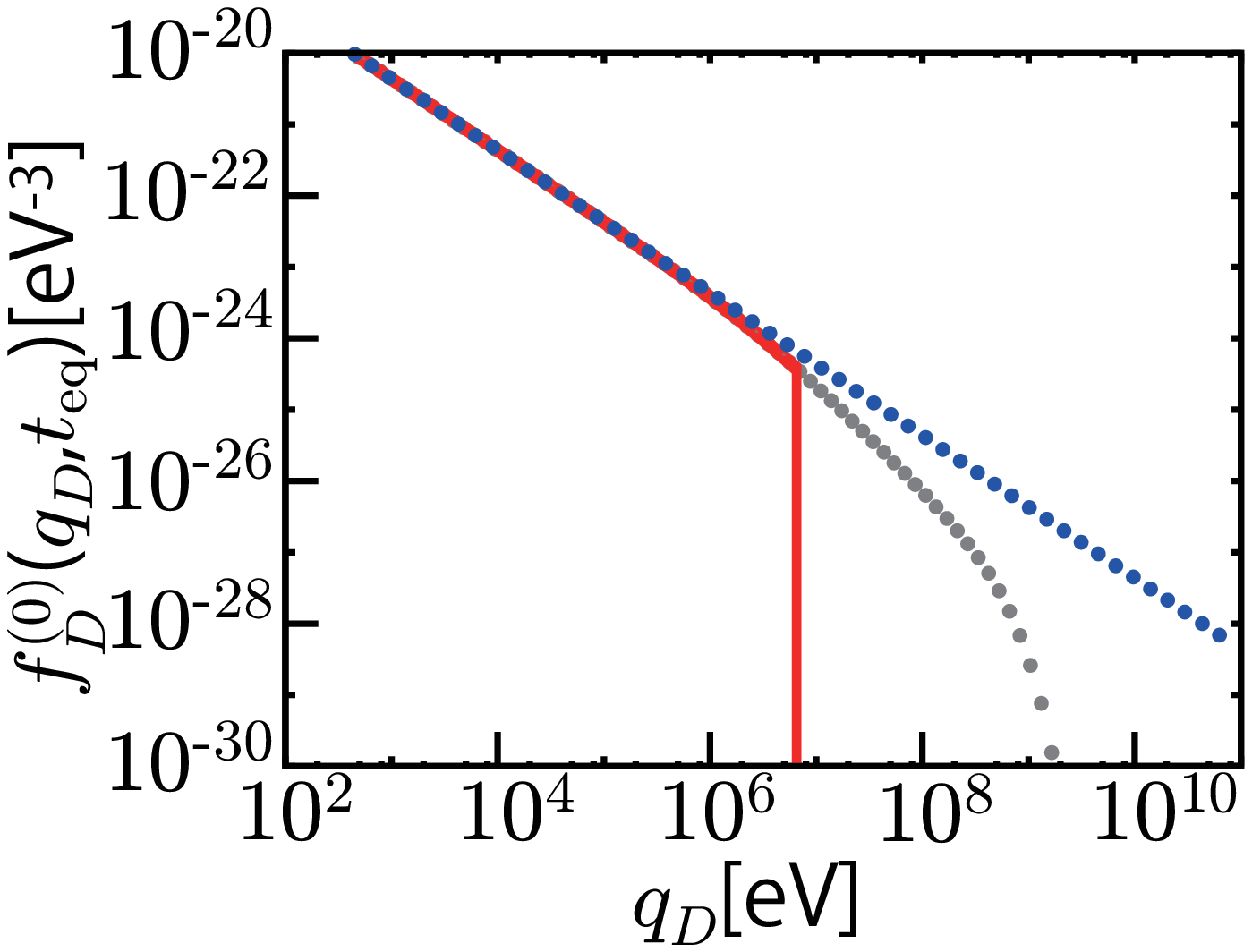}
\includegraphics[width=70mm]{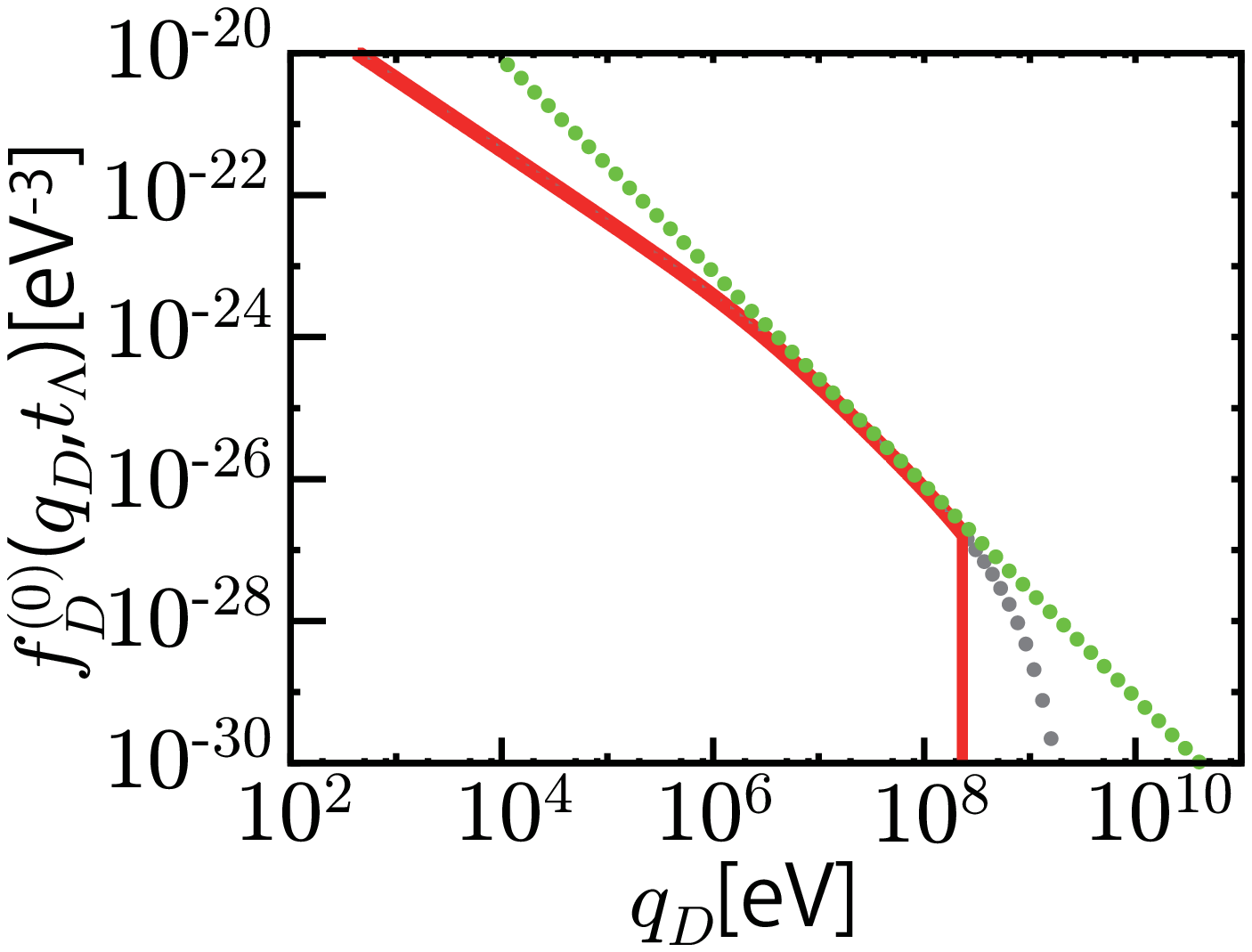} \caption{Snap shots of the
distribution function of daughter particles (red lines) at
matter-radiation equality $t_{\rm eq}$ (left) and matter-lambda equality
$t_\Lambda$ (right). The present distribution function is also given
(gray dotted). The green and blue dotted lines are the fitting lines
with $\propto q_{\rm D}^{-1}$ and $q_{\rm D}^{-3/2}$, which represent
the particles created in the radiation and matter dominated epochs,
respectively. We find that these fitting lines are in good agreement 
with the redlines.}  \label{fig.2}
\end{center}
 \end{figure}

We depict the distribution function of daughter particles at two
different epochs, at matter-radiation and matter-$\Lambda$ equalities,
in figure 2.  We find that the distribution function, $f_{\rm D}$, is
proportional to $q_{\rm D}^{-1}$ and $q_{\rm D}^{-3/2}$ for $q_{\rm D} <
{p_{\rm th}}/{(1+z_{\rm eq})}$ and ${p_{\rm th}}/{(1+z_{\rm eq})}<q_{\rm
D}<{p_{\rm th}}/{(1+z_{\Lambda })}$, respectively, where $z_{\rm eq}$
and $z_{\Lambda }$ are the redshifts of matter-radiation and
matter-$\Lambda $ equalities, respectively. These dependences can be
understood as follows.  First let us estimate the number density of
daughter particles $n$.  When $t\ll \Gamma^{-1}$, the number density of
daughter particles is given as
\begin{equation} 
n_{\rm D}\simeq n_{\rm M} (1-\exp(-\Gamma t))\simeq \dfrac{\Gamma }{H} n_{\rm M}~.
\end{equation}
The number density $n_{\rm D}$ is also expressed through an integration
of $f_{\rm D}^{(0)}$ as,
\begin{equation*} 
n=\displaystyle\int_{0}^{q_{\rm D}}f^{(0)}_{\rm D}(q_{\rm D}^{\prime },t)d^{3}q_{\rm D}^{\prime }\sim q_{\rm D}^{3}f_{\rm D}^{(0)}(q_{\rm D},t)~.\\
\end{equation*}
Thus we can express $f^{(0)}_{\rm D}$ with $H$ and $q_{\rm D}$ as, 
\begin{eqnarray} 
f^{(0)}_{\rm D}&\sim & \dfrac{n_{\rm D}}{q_{\rm D}^{3}}\sim
 \dfrac{\Gamma }{H q_{\rm D}^{3}}n_{\rm M}~.
\end{eqnarray}
In the radiation-dominated epoch, the scale factor $a$ grows
proportional to $t^{1/2}$ and the Hubble parameter $H$ is proportional
to $a^{-2}$.  In addition, the masses of the mother and the daughter
particles determine the momentum of daughter particles at their creation
to a fixed (constant) value $p_{\rm th}$.  Thus $q_{\rm D}$ can be
written as
\begin{equation}
q_{\rm D}\equiv ap_{\rm th}\propto a~.
\end{equation}
Combining the above dependences altogether, we can derive the $q_{\rm
D}$-dependence in figure \ref{fig.2} in the radiation-dominated epoch as
\begin{equation} 
f_{\rm D}^{(0)}(q_{\rm D},t_{\rm eq})\propto q_{\rm D}^{-1}~.
\end{equation}
On the other hand, in the matter-dominated epoch where $a\propto
t^{2/3}$ and $H \propto a^{-3/2}$, we obtain the distribution of
daughter particles created in the matter dominated epoch in the same way
as
\begin{equation} 
f_{\rm D}^{(0)}(q_{\rm D},t_{\Lambda})\propto q_{\rm D}^{-3/2}~.
\end{equation}
We confirm that these dependencies are indeed found in figure \ref{fig.2}.

The energy density of daughter particles can be calculated in the
same way as mother's. It is given by
\begin{eqnarray}  
\rho_{\rm D}&=&\dfrac{\rho_{\rm c}\Omega_{\rm DM}}{m_{\rm M}a^{4}}\displaystyle\int_{0}^{+\infty } 4 \pi q_{\rm D}^{2}\sqrt{m_{\rm D1}^{2}a^{2}+q_{\rm D}^{2}}f_{\rm D}^{(0)}(q_{\rm D},t)dq_{\rm D} \\ \nonumber
& &+\dfrac{\rho_{\rm c}\Omega_{\rm DM}}{m_{\rm M}a^{4}}\displaystyle\int_{0}^{+\infty } 4 \pi q_{\rm D}^{2}\sqrt{m_{\rm D2}^{2}a^{2}+q_{\rm D}^{2}}f_{\rm D}^{(0)}(q_{\rm D},t) dq_{\rm D} ~.
\end{eqnarray}

Finally the time evolution of an homogeneous and isotropic expanding
universe follows the Friedmann equation,
\begin{equation} 
\dfrac{\dot{a}}{a}=\sqrt{\dfrac{8\pi G}{3}(\rho_{\rm M}+\rho_{\rm
 D}+\rho_{\rm B}+\rho_{\gamma}+\rho_{\nu}+\rho_{\Lambda})}~,
\end{equation}
where $\rho_{\rm B}$, $\rho_{\gamma}$, $\rho_\nu$ and $\rho_{\Lambda}$ are the
densities of baryon, photon, neutrinos, and dark energy, respectively.

\section{Result and Discussion}
\subsection{Background energy densities}
\begin{figure}[htbp]
 \begin{center}
  \includegraphics[width=100mm]{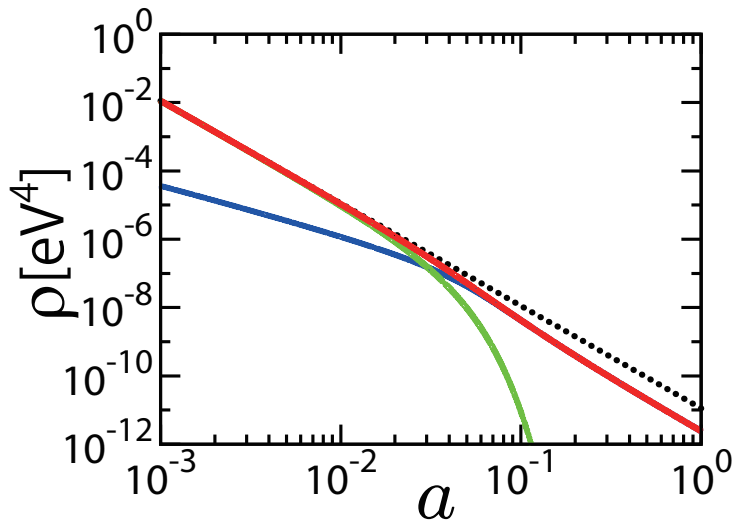}
 \end{center}
 \caption{Evolution of energy densities of the mother (green) and
 daughter (blue) particles
 as a function of scale factor $a$ with model parameters $m_{\rm M}=1.0$
 TeV, $m_{\rm D}=0.20$ TeV, and $\Gamma^{-1}=0.1$ Gyr. The total energy
 density of mother and daughter particles is shown as a red line. The time evolution for the standard CDM
 model is also shown for comparison (black dotted line).}
 \label{fig.3}
 \end{figure}

By integrating the distribution functions $f_{\rm M}^{(0)}$ and $f_{\rm
D}^{(0)} $, we obtain the time evolution of energy densities of 
mother and daughter particles, which is shown in figure
\ref{fig.3}.
As is shown in the figure, if the dark matter particles decay, the total energy density in the
universe becomes small comparedwith the standard $\Lambda$-CDM
model \cite{2000astro.ph..2400Z,2004PhRvD..69f3512Z}. Thus the time evolution of the scale factor $a$ differs from that of
the $\Lambda $-CDM model. This leads to the different angular diameter
distances to the last scattering surface of CMB $d_{\rm A}(z_{*})$ and
the position of BAO $d_{z}$. The distances to CMB and the position of
BAO are measured precisely by WMAP \cite{2011ApJS..192...14J} and SDSS
\cite{2010MNRAS.401.2148P}, respectively. The uncertainties of these
measurements are also available in those papers which can be used to
constrain the decay rate $\Gamma $ and the mass ratio $m_{\rm D1}/m_{\rm
M}$, as we will discuss below.

\subsection{Constraints from Hubble constant, BAO and CMB}
In this paper we only consider the dark matter which decays after
cosmological recombination. For this reason, we fix standard
cosmological parameters to the values which agree with the $\Lambda
$-CDM model in the early universe obtained by WMAP7
\cite{2011ApJS..192...14J}.  Because the decay of mother particles can
be neglected deep in the radiation dominated era, the initial conditions
of the dark matter energy density $\rho_{\rm M}$ and the scale factor
$a(t_{\rm i})$ can be set as in the $\Lambda$-CDM model without decay.
In this paper we use the relation which holds in the radiation dominated
era, $a(t=0.02$ sec $)= 4.60 \times 10^{-12}$ as our initial condition.

The "comoving" angular diameter distance to CMB, $d_{\rm A}(z_{*})$,
 is sensitive to the deviation of $a$, where $z_{*}$ is the shift of
recombination. Here $d_{\rm A}(z_{*})$ is determined precisely by WMAP
project as
\begin{equation*} 
d_{\rm A}(z_{*})=14116^{+160}_{-163}~{\rm Mpc}~.
\end{equation*}

In the same sense the distance to BAO deviates from that in the
$\Lambda$-CDM model. According to the observation of BAO by SDSS
\cite{2010MNRAS.401.2148P}, the BAO constraint is given through the
variable defined as $d_{z}\equiv r_{s}(z_{\rm D})/D_{V}(z) $, where
$r_{s}(z_{\rm D})$ is the comoving sound horizon at the baryon drag
epoch and $z_{\rm D}$ is the redshift when photons decouple baryons.
According to the report of WMAP7 \cite{2011ApJS..192...14J},
\begin{equation} 
r_{s}(z_{\rm D})=153.2\pm 1.7~\text{Mpc}~.
\end{equation} 
The distance $D_{V}(z)$ is a function of redshift $z$ defined as \cite{2005ApJ...633..560E,2007MNRAS.381.1053P},
\begin{equation} 
D_{V}(z)\equiv \left( \dfrac{(1+z)^{2}zd_{\rm A}^{2}(z)}{H(z)}\right)^{1/3},
\end{equation}
where $d_{\rm A}(z)$ is the angular diameter distance to the point whose
redshift is $z$. When one specifies a
cosmological model and the evolution of the scale factor $a$ is determined,
we can use $d_{z}$ to constraint the parameters of the model.  The
distances $d_{0.2}$ and $d_{0.35}$ are given by SDSS
DR7 as \cite{2010MNRAS.401.2148P},
\begin{eqnarray} 
d^{\rm obs}_{0.20}&=&0.1905\pm 0.0061~, \\
d^{\rm obs}_{0.35}&=&0.1097\pm 0.0036~.
\end{eqnarray}
We define a vector $\boldsymbol{x}$ as
\begin{equation}
\boldsymbol{x}=\left( 
\begin{matrix}
d^{\rm th}_{0.20}-{d}^{\rm obs}_{0.20} \\
d^{\rm th}_{0.35}-{d}^{\rm obs}_{0.35} \\
\end{matrix}
\right)~,
\end{equation}
where $d^{\rm th}_{0.20},d^{\rm th}_{0.35}$ are the distances based on
the cosmological model to be constrained.  The matrix $C^{-1}$, which is
inverse of the covariance matrix $C\equiv
\left<\boldsymbol{x}\,^t\boldsymbol{x}\right>$, is given by,
\begin{equation} 
C^{-1}=\left(
\begin{matrix}
30124 & -17227 \\
-17227 & 86977 \\
\end{matrix}
 \right)~,
\end{equation}
where $^t\boldsymbol{x}$ is transpose of the vector $\boldsymbol{x}$.
We use the value $\chi^{2}\equiv~^t\boldsymbol{x}C^{-1}\boldsymbol{x}$ for our chi-square test. Since we have two model
parameters, the regions of $1\sigma $ and $2 \sigma $ confidence levels
correspond to those which satisfy $\Delta \chi^{2}<2.18$ and $6.30$ from
the minimum, respectively.

\begin{figure}[htbp]
 \begin{center}
  \includegraphics[width=100mm]{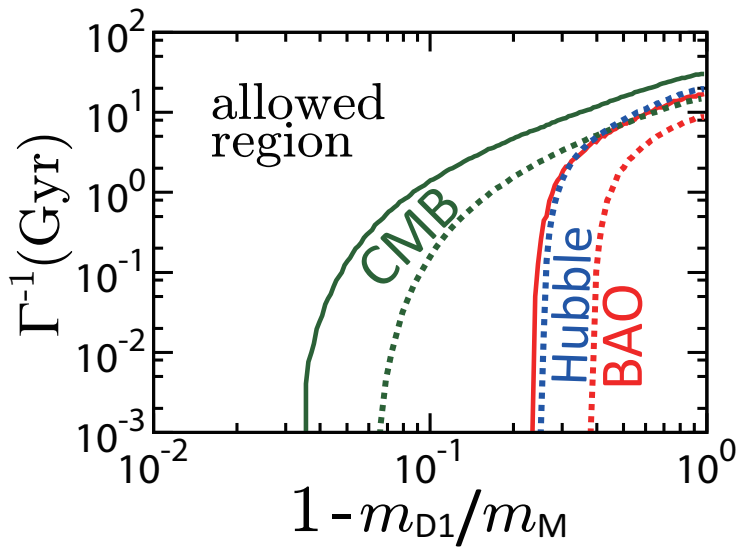}
 \end{center}
 \caption{Constraints on the lifetime of decaying dark matter and
 the mass ratio between the mother and the daughter particles. Solid
 lines and dashed lines correspond to 1 $\sigma $ C.L. and 2$\sigma $
 C.L. of the constraints, respectively. The names of observational data
 are also shown in the figure. }
 \label{fig.4}
 \end{figure}

Due to the dark matter decay, the total energy density in our model is
always lower than that in the $\Lambda $-CDM model if the energy
densities of dark matter in the early universe are fixed to the same
value between the two models. Thus the Hubble parameter $H$ of our model
is always smaller than that in the $\Lambda $-CDM, which we denote as
$H_{\Lambda - \rm CDM}$.
 A simple constraint on our model is therefore put from the current
value of the Hubble parameter. The latest compilation determines the
Hubble constant as $H_0=73.8\pm 2.4$ km/s/Mpc \cite{2011ApJ...730..119R}
including systematics. By comparing the Hubble parameter with the data,
we obtain constraints on the lifetime of the mother particle
$\Gamma^{-1}$ and the mass ratio $m_{\rm D1}/m_{\rm M}$, which are shown in figure \ref{fig.4}.

We consider the "comoving" angular diameter distance toward the last scattering
surface $d_{\rm A}(z_{*})$, where $z_{*}$ is the redshift of
recombination. The distance $d_{\rm A}$ can be written
as
\begin{equation} 
d_{\rm
 A}=\displaystyle\int_{0}^{z_{*}}\dfrac{dz}{H(z)}>\displaystyle\int_{0}^{z_{*}}\dfrac{dz}{H_{\Lambda -{\rm CDM}}(z)}~.
\end{equation}
Therefore 
\begin{equation} 
d_{\rm A}(z_{*})>d_{\rm A(\Lambda-{\rm CDM})}(z_{*})~.
\end{equation}

In figure \ref{fig.4} we show the constraint on the lifetime of decaying
dark matter from the distance to BAO and the last scattering surface of
CMB.  The $\Lambda -$CDM model corresponds to the limits $\Gamma \to 0 $
and/or $\left(1-{m_{\rm D1}}/{m_{\rm M}}\right)\to 0 $. As $\Gamma$ becomes
smaller, or $m_{\rm D1}$ approaches closer to $m_{\rm M}$, the distance
$d_{\rm A}(z_{*})$ approaches to the value in the $\Lambda $-CDM
model. In general, constraints from CMB are stronger than those from
BAO. This is simply because CMB data are more precise than the current
data of BAO. In the massless limit of daughter particles, we find
$\Gamma^{-1} >30$ Gyr from CMB (at 1$\sigma $). Because we have derived
the constraint from the background $f_{\rm D}^{(0)}(q_{\rm D},t)$ only,
this constraint is weaker than that obtained by
\cite{2004PhRvL..93g1302I,2009JCAP...06..005D} in which the
perturbations are included. At the limit $\Gamma^{-1}\to 0$ we obtain
$m_{\rm D1}>0.97 m_{\rm M}$ from CMB (at 1$\sigma $). These values are
consistent with the results obtained by the simplified estimations without
using the distribution functions (see Appendix a).

\subsection{Free-streaming scale}
The daughter particles created by the decay of mother particles move
with large velocity in the expanding universe. It leads that the depth
of the gravitational potential of mother particles becoming shallower which have been created by a group of 
mother particles becomes shallower in time. In the linear theory
of structure formation, the structures smaller than the free-streaming
scale $l_{\rm FS}$ are erased. Since it is difficult to fully calculate
the density perturbations of the mother and the daughter particles, here we use
$l_{\rm FS}$ for constraints on $\Gamma^{-1}$ and $m_{\rm D1}/m_{\rm M}$.
We restrict our attention to the case where $\Gamma ^{-1}$ is much smaller
than the age of universe at redshift $z$, $\Gamma^{-1} \ll t \sim H^{-1}(z)$. In this
case, because almost all the mother particles have decayed by that time
into daughter particles which are responsible for the depth of gravitational
potential of mother particles becoming very shallow, we can place constraints from a simple argument that the
scale of any observed structure bounded by dark matter should be larger
than the free-streaming scale of daughter particles.

Free-streaming scale $l_{\rm FS}(z)$ at each redshift $z$ can be
estimated using the averaged velocity of daughter particles
$\overline{v}(z)$ as
\begin{equation} 
l_{\rm FS}(z) \sim \overline{v}(z) \times \dfrac{1}{H(z)}~.\label{5.3}
\end{equation}
Here we define $\overline{v}(z)$ as
\begin{equation*} 
\overline{v}(z) \equiv  \dfrac{\displaystyle\int_{0}^{p_{\rm th}}4\pi q_{\rm D}^{2} v(q_{\rm D},z)f_{\rm D}^{(0)}dq_{\rm D} }{\displaystyle\int_{0}^{p_{\rm th}}4\pi q_{\rm D}^{2} f_{\rm D}^{(0)}dq_{\rm D} }~,
\end{equation*}
where
\begin{equation*} 
v(q_{\rm D},z)\equiv \dfrac{q_{\rm D}}{\sqrt{q_{\rm D}^{2}+m_{\rm D1}^{2}a^{2}}}~.
\end{equation*}
Here $v(q_{\rm D},z)$ is the magnitude of physical velocity of the daughter particle
whose comoving momentum is $q_{\rm D}$ at redshift $z$.

\begin{figure}
 \begin{center}
  \includegraphics[width=70mm]{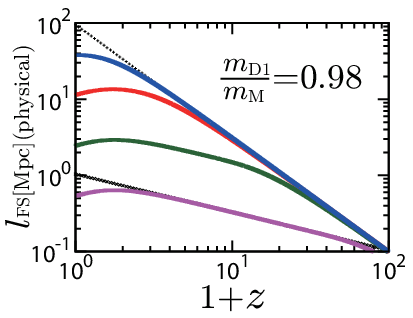}
  \includegraphics[width=70mm]{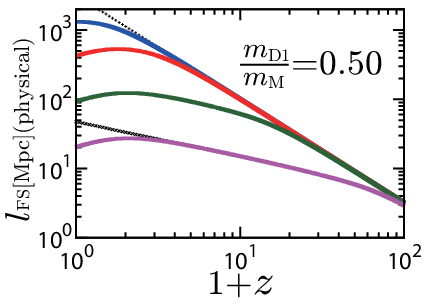}
 \end{center}
  \caption{Time evolution of the free-streaming scale of daughter
 particles for ${m_{\rm D1}}/{m_{\rm M}}=0.98$ (left) and $0.50$
 (right). The different lines in the panels correspond to the different
 lifetime of the mother particle: $\Gamma =0.01$ Gyr (magenta line),
 $0.1$ Gyr (green line), $1.0$ Gyr (red line), and $10.0$ Gyr (blue
 line), respectively. Black dotted lines with steeper slope are the
 fitting lines to the case with particles having a constant momentum,
 which are proportional to $(1+z)^{-3/2}$, and the others are to the
 case with non-relativistic particle, which is proportional to
 $(1+z)^{-1/2}$. See main text for details.}  \label{fig.5}
  \end{figure}
\begin{figure}[htbp]
 \begin{center}
  \includegraphics[width=100mm]{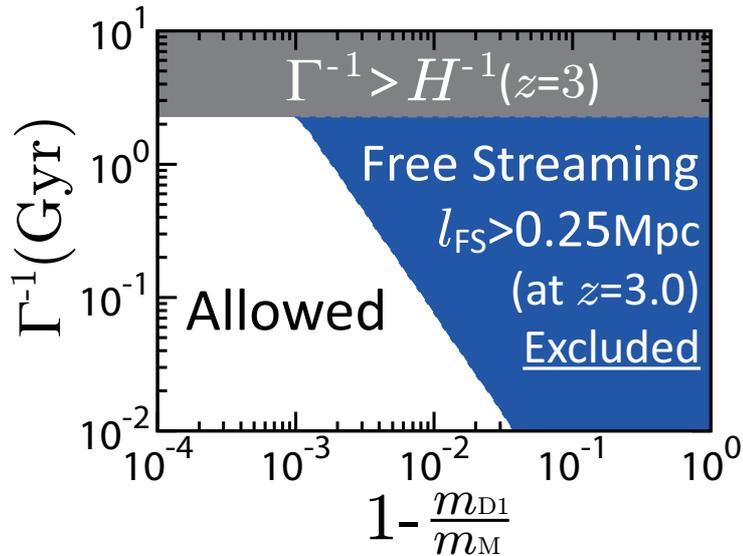}
 \end{center}
 \caption{Constraint on the lifetime of decaying dark matter from the
 free-streaming scale. The range painted over in gray corresponds to
 $\Gamma^{-1} > H^{-1}(z=3)$, at which the gravitational potential from the mother
 particles would be significant at $z=3$, and hence it cannot be excluded.}  \label{fig.7}
 \end{figure}

We calculate the free-streaming scale $l_{\rm FS}$ of daughter particles
at each redshift $z$, which is shown in figure \ref{fig.5}.  We explain
these curves in the matter dominated epoch ($1\lesssim z\lesssim 3000$)
as follows.  First, when $\Gamma^{-1}>H^{-1}$ the daughter particles
with a constant physical momentum $p_{\rm th}$ are kept being created
and the averaged physical velocity of daughter particles becomes
constant. Thus
\begin{equation}
l_{\rm FS}\sim \dfrac{Const.}{H(z)}\propto (1+z)^{-3/2}~,
\end{equation}
where we have used the fact that $H(z)$ is proportional to $a^{-3/2} =
(1+z)^{3/2}$ in the matter dominated epoch.   

Secondly, when $\Gamma^{-1}<H^{-1}$, i.e., the decay process has
finished, the averaged velocity of daughter particles decays as
$\overline{v} \propto a^{-1}=(1+z)$. Therefore
\begin{equation}
l_{\rm FS}\propto \dfrac{(1+z)}{H(z)}\propto (1+z)^{-1/2}~.
\end{equation}
One can see that these dependences well describe the calculated curves in
the figure.

For the dark matter to form structures such as dark halos, the
free-streaming scale $l_{\rm FS} $ should be less than the size of the
structures.  We find in figure \ref{fig.5} that the free-streaming scale
of daughter particles is sensitive to the lifetime $\Gamma^{-1}$ and
$m_{{\rm D}1}/m_{{\rm M}}$. Contrary to the case with massless
daughter particles, the free-streaming scale of daughter particles
becomes smaller if the lifetime of the mother particle becomes shorter. The
reason is that the velocity of massive daughter particles decays faster
in the earlier universe, because the expansion of the universe is
faster. Hence the velocity of daughter particles decays in shorter
timescale if they decay earlier, which leads to the smaller free
streaming scale.  As expected, the free-streaming scale becomes smaller
if the mass ratio $m_{{\rm D}1}/m_{{\rm M}}$ becomes smaller.  Therefore
any existence of large scale structure by dark matter can be used to
constrain $\Gamma^{-1}$ and $m_{{\rm D}1}/m_{{\rm M}}$ through the free
streaming scale.

From the observations of Lyman $\alpha$ cloud at $z\lesssim 3$, the
density fluctuations at about 1 Mpc comoving scale have been found, for
example, in SDSS
\cite{1996ApJ...473..576F,2009JCAP...05..012B}. Therefore, when
$\Gamma^{-1} \ll H^{-1}(z= 3)$ is satisfied, the range of mass ratio $m_{\rm
D1}/m_{\rm M}$ is excluded if $l_{\rm FS}\ge 1 $ Mpc (0.25 Mpc in
physical scale) at $z = 3 $.  By taking this into account, we obtain the
constraint on the lifetime of the mother particle as shown in figure
\ref{fig.7}.  We find that the free streaming scale has a constraining
power even for $\left(1-m_{\rm D1}/{m_M}\right)\lesssim 0.01$, and the
constraint is complementary to those obtained from
the geometric distances to CMB and BAO.
In the case $\Gamma^{-1} \gtrsim H^{-1}(z)$, on the other hand, we can
not use this method since most of the mother particles, and therefore
the gravitational potentials, still remain at the redshift $z$. A full
treatment of cosmological density perturbations will be necessary in this case.
Because our Lyman $\alpha$ constraint is given at $z=3$, the region
corresponding to $\Gamma^{-1} \gtrsim H^{-1}(z=3)$ (grey region in figure
\ref{fig.6}) can not be excluded from the free streaming scale for now.

\subsection{Comparison with Peter et al.}

\begin{figure}[htbp]
 \begin{center}
  \includegraphics[width=100mm]{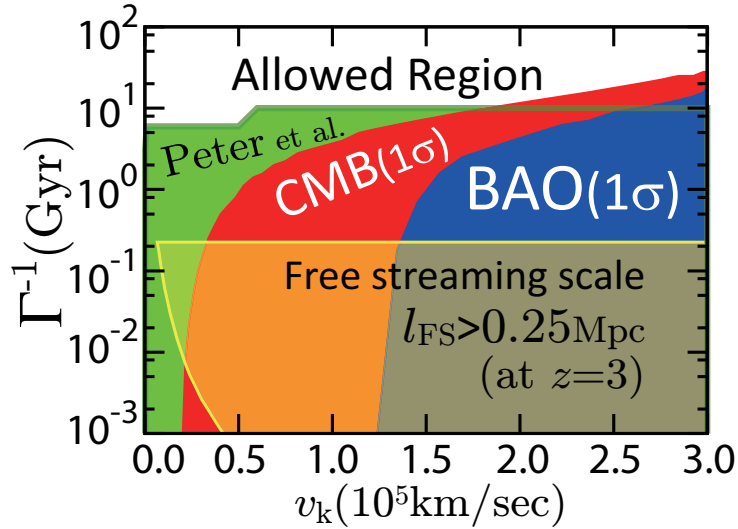}
 \end{center}
 \caption{Comparison with the results from Peter
 et al. \cite{2010PhRvD..81h3511P} on the life time and the kick
 velocity plane. Red, blue and yellow regions
 are the parameter space that we exclude in this paper.}
 \label{fig.6}
 \end{figure}
 
Before ending this section we compare our result with the constraints
obtained by Peter et al. \cite{2010PhRvD..81j3501P}. They considered a
dark matter decay and describe it in terms of the velocity of 
daughter particles $v_{\rm k}$ at their creation (they call it "kick
velocity"). In a virialized dark halo, matter moves slower than its
associated virial velocity $v_{\rm vir}$ \cite{2003ApJ...591L.107S}. When $v_{\rm k}>v_{\rm vir}$,
dark-matter halos will be disrupted by these particle decays. They
performed several numerical simulations in order to study the detailed
evolution of the total mass and density profile of galaxies composed of
particles that undergo such velocity kicks as a function of the kick
speed. As a result, $v_{\rm k}$ is strictly restricted from the
stability of the halos. We find that our constraints are 
comparable with their results in high kick velocity region, as shown in
figure \ref{fig.6}. Note that our constraints are completely independent
from theirs and we believe that our constraints are less uncertain in
that they are free from the variety of galaxies.

\section{Conclusion}
In this paper we consider a decaying dark matter model in which
the  massive mother particle decays into two massive and massless daughter
 particles after cosmological recombination. We derive a complete set of Boltzmann equations to describe the evolution of the particles. 
We obtain constraints on the lifetime of the mother particle
$\Gamma^{-1} $ and the mass
 ratio ${m_{\rm D1}}/{m_{\rm M}}$ with $m_{\rm D2}=0$ 
from the Hubble parameter, CMB and BAO. The allowed $\Gamma ^{-1}$ decreases monotonically as $m_{\rm
 D1}/m_{\rm M}$ increases. For the free streaming constraint, on the other hand, we find the opposite dependence. We find $\Gamma^{-1} >30$ Gyr at the massless
 limit of daughter particles and $m_{\rm D1}>0.97 m_{\rm M}$ at the
 limit $\Gamma^{-1}\to 0$, from the distance to CMB (1$\sigma $).
 We also obtain constraints from the free-streaming of daughter
 particles from observations of Lyman $\alpha$ as $\left(\Gamma^{-1}/10^{-2}{\rm Gyr}\right) \lesssim \left((1-m_{\rm D1}/m_{\rm M})/10^{-2}\right)^{-3/2}$ for $\Gamma^{-1} < H^{-1}(z=3)$. 
However, to extend the free-streaming constraint to the range
$\Gamma^{-1}\gtrsim H^{-1}$ or to include the information from density
perturbations such as CMB angular power spectrum, a complicated
calculation of density perturbations is
 necessary. This will be an interesting subject and presented in a
 separate paper.

\acknowledgments This work has been supported in part by Grant-in-Aid
for Scientific Research Nos. 22012004 (K.I.), 22340056 and 18072004
(N.S.) of the Ministry of Education, Sports, Science and Technology
(MEXT) of Japan, and also supported by Grant-in-Aid for the Global
Center of Excellence program at Nagoya University "Quest for Fundamental
Principles in the Universe: from Particles to the Solar System and the
Cosmos" from the MEXT of Japan. This research has also been supported in part by World Premier International
Research Center Initiative, MEXT, Japan.

\section*{Appendix a Massless limit of daughter particle}
In the case that a mother particle decays into two massless
particles, the energy densities of mother particle $\rho_{\rm M}$ and
daughter radiation $\rho_{\rm D}$ satisfy the following
equations 
\begin{equation*} 
\dot{\rho }_{\rm M}=-3H\rho_{\rm M}-\Gamma \rho_{\rm M}~,
\end{equation*}
and
\begin{equation*} 
\dot{\rho }_{\rm D}=-4H\rho_{\rm D}+\Gamma \rho_{\rm M}~.
\end{equation*}
These differential equations can be solved to give
\begin{equation*}
\rho_{\rm M}=\rho_{\rm M \emptyset }a^{-3}\exp (-\Gamma t)~,\\
\end{equation*}
and
\begin{equation*}
\rho_{\rm D}=\dfrac{\Gamma \rho_{\rm M\emptyset }}{a^{4}}\displaystyle\int_{0}^{t}a(t^{\prime })\exp (-\Gamma t^{\prime } )dt^{\prime },
\end{equation*}
where $\rho_{{\rm M}\emptyset }$ is the expected energy density of
dark matter without decay. 
In this calculation we obtained the following constraints,
\begin{equation*} 
\Gamma^{-1}\ge 30 {\rm Gyr}~,
\end{equation*}
and
\begin{equation*} 
\Gamma^{-1}\ge 18 {\rm Gyr}~,
\end{equation*}
from the angular diameter distances to CMB and the position of BAO,
respectively. These values agree with the results in the massless limit obtained from calculations in
the main text, where the distribution functions of daughter particles are
directly integrated instead using the above simple equations.

Next we show how one can obtain a limiting value of the mass ratio
$m_{\rm D1}/m_{\rm M}$ in the limit $\Gamma^{-1}\to 0 $. In this limit
our model corresponds to the $\Lambda $-CDM model such that $\Omega_{\rm
M}$ is replaced with $\dfrac{m_{\rm D1}}{m_{\rm M}}\Omega_{{\rm
M}\emptyset}$ where $\Omega_{{\rm M}\emptyset}$ is the expected
density parameter of dark matter without decay. We can deduce a
constraint on the ratio $m_{\rm D1}/m_{\rm M}$ from a constraint on
$\Omega_{\rm M}$ from the angular diameter distances to CMB and BAO.
In this way we obtain the following constraints, 
\begin{equation*} 
m_{\rm D1}/m_{\rm M}\ge 0.97,
\end{equation*}
and
\begin{equation*}
m_{\rm D1}/m_{\rm M}\ge 0.23,
\end{equation*}
from the angular diameter distances to CMB and BAO, respectively. 
These values are roughly in agreement with the values in figure \ref{fig.4}, in
which the distribution functions of daughter particles are
directly integrated.

\section*{Appendix b Present tempreture of CDM mother particle}

The temperature of mother particles at present time $T_{\rm M0} $ can be
estimated by taking into account of the decoupling temperature $T_{\rm
d}$ of mother particles \cite{1990eaun.book.....K}.  The
temperature which corresponds to the mass of the mother particle $m_{\rm
M} = 1$ TeV is so high that all species of standard particles such as
eight gluons, $W^{\pm }$, $Z^0$, three generations of quarks and
leptons, and one complex Higgs doublet are relativistic at the
decoupling of mother particles. We define $g_{*}$ as the total number of
effectively massless degrees of freedom. In the epoch that the mother
particles are in thermal equilibrium, we expect $g_{* \rm
(early)}=106.75$, while $g_{* \rm(now)}=3.36$ at present.

The comoving number density of photons when the mother particles are in
thermal equilibrium is
\begin{equation} 
n_{\gamma }=\dfrac{g_{*(\rm now)}}{g_{*(\rm early)}}n_{\gamma 0}~.
\end{equation}
Here the number density of photons at present $n_{\gamma 0}$ is 
$410$ cm$^{-3}$
\cite{1996ApJ...473..576F,1990eaun.book.....K} and therefore
\begin{equation*} 
n_{\gamma }=12.9~\text{cm}^{-3}~.
\end{equation*}
On the other hand, the baryon number density can be derived from the
baryon-photon ratio today $\eta = n_{\rm b}/n_{\gamma }$, which is
estimated as $\eta=(6.19~\pm 0.15 )\times 10^{-10}$ ($1 \sigma $ C.L.)
\cite{2011ApJS..192...18K}.  The ratio of the number densities between
CDM mother particles and photons today, $n_{\rm M}/n_{\gamma 0}$, can be
written as
\begin{equation} 
\dfrac{n_{\rm M}}{n_{\gamma 0}}=\dfrac{n_{\rm M}}{n_{\rm b}}\dfrac{n_{\rm b}}{n_{\gamma 0}}=\eta \times \dfrac{\Omega_{\rm DM}}{\Omega_{\rm b} }\dfrac{m_{\rm p}}{m_{\rm M} }~, \label{eq:3.7}
\end{equation}
where $\Omega_{\rm b} $ is the cosmological density parameter of
baryon. Here we have neglected a contribution from helium.  By
substituting $m_{\rm M}=1.0$ TeV, $\Omega_{\rm M} =0.222$, and
$\Omega_{\rm b}=0.0446$ into eq.(\ref{eq:3.7}), we obtain
\begin{equation*} 
\dfrac{n_{\rm M}}{n_{\gamma 0}}=2.89\times 10^{-12}~.
\end{equation*}
The time of the thermal decoupling of mother particles can be estimated
through the relation of the Boltzmann factor as, 
\begin{equation*} 
\dfrac{n_{\rm M}}{n_{\gamma }}=\exp \left( -\dfrac{m_{\rm M}}{T_{\rm d}} \right)~.
\end{equation*}
Thus we obtain the temperature of mother particles as
\begin{equation} 
T_{\rm M0}= a_{\rm d}^{2}T_{\rm d} \simeq 1.7\times 10^{-14}~\text{K}=1.4\times 10^{-18}~\text{eV}{}~,
\end{equation}
where $a_{\rm d}$ is the scale factor when the mother particles
decouple from the thermal bath. We can write $a_{\rm d}$ as
\begin{equation*} 
a_{\rm d}\sim \dfrac{T_{\gamma 0}}{T_{\rm d}},
\end{equation*}
where $T_{\gamma 0}=2.725$ K is the temperature of CMB \cite{1996ApJ...473..576F}.

\section*{Appendix c First order Boltzmann equations for daughter particle}
In this appendix we give a set of Boltzmann equations at first order
which are necessary to compute the evolution of density perturbations
associated with the mother and daugher particles.
The standard linear theory of density perturbations has
been presented, for example, in the synchronous and the conformal
Newtonian gauges \cite{1995ApJ...455....7M}. Here we expand
the linear perturbation theory by taking into account the decays of a
SUSY CDM particle into two daughter particles
\cite{1999PhRvD..60l3508K}.  

By substituting eq.(\ref{eq:2.10}) and (\ref{eq:2.11}) into
eq.(\ref{eq:2.1}) and (\ref{eq:2.3}), respectively, and comparing the
equations order by order, we obtain the following equations at first th order
\begin{equation}
\text{1st order}:\dot\Psi_{\rm M}+i\dfrac{q_{\rm M}}{a\epsilon_{\rm M} }({\boldsymbol{k}}\cdot \hat{\boldsymbol{n}})\Psi_{\rm M} +\dfrac{d \ln{f^{(0)}_{\rm M}}}{d \ln{q_{\rm M}}}\left(\dot{\eta}-\dfrac{1}{2}(\dot{h}+6\dot{\eta})(\hat{\boldsymbol{k}}\cdot \hat{\boldsymbol{n}})^{2}\right)=0~. \label{2.13}
\end{equation}
Here $\epsilon_{\rm M}=\sqrt{q_{\rm M}^2+a^2 m_{\rm M}^2}$ is the energy of mother particles, and $h$ and
$\eta $ are the metric perturbations in the synchronous gauge defined
from the perturbed (space-space) metric $h_{ij}$ as
\begin{equation*} 
h_{ij}=\displaystyle\int \left( \hat{\boldsymbol{k}}_{i} \hat{\boldsymbol{k}}_{j}h(\boldsymbol{\boldsymbol{k}},t) +(\hat{\boldsymbol{k}}_{i} \hat{\boldsymbol{k}}_{j} - \dfrac{1}{3}\delta_{ij})6\eta (\boldsymbol{\boldsymbol{k}},t) \right) \exp ( -i \boldsymbol{k}\cdot \boldsymbol{x}) d^{3}\boldsymbol{k}~,
\end{equation*}
where $\boldsymbol{k}\equiv k \hat{\boldsymbol{k}}$ is a wave number
vector.  Hereafter, we omit the arguments of
$h(\boldsymbol{\boldsymbol{k}},t)$ and $\eta
(\boldsymbol{\boldsymbol{k}},t)$ for simplicity. For daughter particles,
we obtain
\begin{eqnarray}
\text{1st order} &:&\dot{\Psi }_{{\rm D}j}
+i\dfrac{q_{\rm D}}{a\epsilon_{{\rm D}j} }
(\boldsymbol{k}\cdot \hat{\boldsymbol{n}})\Psi _{{\rm D}j}
+\dfrac{d \ln{f_{\rm D}^{(0)}}}{d \ln{q_{\rm D}}}\bigg(\dot{\eta}-\dfrac{1}{2}(\dot{h}+6\dot{\eta})(\hat{\boldsymbol{k}}\cdot \hat{\boldsymbol{n}})^2\bigg) \nonumber \\
& &=\frac{1}{f_{{\rm D}j}^{(0)}}\displaystyle\int \Gamma (q_{\rm D},q_{\rm M})f_{\rm M}^{(0)}(q_{\rm M})(\Psi_{\rm M}(q_{\rm M})-\Psi_{{\rm D}j}(q_{\rm D}))d^{3}q_{\rm M} \nonumber \\
& &=\dfrac{4\Gamma }{B_{\rm M}}\dfrac{q_{\rm D}^{\prime }}{q_{\rm
 D}}\frac{f_{\rm M}^{(0)}(q^{\prime }_{\rm D})}{f_{{\rm D}j}^{(0)}(q_{\rm D})} \left( \Psi_{\rm M}(q^{\prime }_{\rm D})-\Psi _{{\rm D}j}(q_{\rm D}) \right) ~,\label{2.15}
\end{eqnarray}
{Note that on the RHS of the above equation, while the first term
directly comes from
the collision term, the second term results from the change of the
background evolution of the distribution function, i.e., $\dot{f}_{{\rm D}j}^{(0)}$.}

Following \cite{1995ApJ...455....7M}, we consider the Legendre expansion of $\Psi_{\rm M}$ and $ \Psi_{\rm D}$ into $\Psi_{{\rm M}l}$ and $\Psi_{{\rm D}l}$,
respectively. Here the coefficients $\Psi_{{\rm M}l}$ and $\Psi_{{\rm
D}l}$ are defined as, respectively,
\begin{equation}
\Psi_{\rm M} (\boldsymbol{k},\hat{\boldsymbol{n}},q,t)\equiv \displaystyle\sum_{l=0}^{+\infty}(-i)^{l}(2l+1)\Psi_{{\rm M}l}(\boldsymbol{k},q,t)P_{l}(\hat{\boldsymbol{k}}\cdot \hat{\boldsymbol{n}})~, 
\end{equation}
and
\begin{equation}
\Psi_{{\rm D}j} (\boldsymbol{k},\hat{\boldsymbol{n}},q,t)\equiv \displaystyle\sum_{l=0}^{+\infty}(-i)^{l}(2l+1)\Psi_{{\rm D}jl}(\boldsymbol{k},q,t)P_{l}(\hat{\boldsymbol{k}}\cdot \hat{\boldsymbol{n}})~,
\end{equation}
where $P_{l}(\hat{\boldsymbol{k}}\cdot \hat{\boldsymbol{n}})$ are a series of
Legendre polynomials. The factor $(-i)^{l}(2l+1)$ is chosen to simplify
the expansion of plane wave. Note that equation eq.(\ref{2.13}) is the
same as the case of massive particles such as neutrinos without decay
\cite{1995ApJ...455....7M}. Therefore we obtain the recursion formula
for the mother particles without any explicit $\Gamma$ dependence as
\cite{1995ApJ...455....7M},
\begin{eqnarray*} 
\dot{\Psi}_{\rm M0}&=&-\dfrac{q_{\rm M}k}{a\epsilon_{\rm M} }\Psi_{\rm M1}+\dfrac{1}{6}\dot{h}\dfrac{d\ln{f_{\rm M}^{(0)}}}{d\ln{q_{\rm M}}}~,\\
\dot{\Psi}_{\rm M1}&=&\dfrac{q_{\rm M}k}{3a\epsilon_{\rm M} }(\Psi_{\rm M0}- 2\Psi_{\rm M2})~,\\
\dot{\Psi}_{\rm M2}&=&\dfrac{q_{\rm M}k}{5a\epsilon_{\rm M} }(2\Psi_{\rm M1}- 3\Psi_{\rm M3})-\bigg( \dfrac{1}{15}\dot{h}+\dfrac{2}{5}\dot{\eta }\bigg)\dfrac{d\ln{f_{\rm M}^{(0)}}}{d\ln{q_{\rm M}}}~,\\
\dot{\Psi}_{{\rm M}(n)}&=&\dfrac{q_{\rm M}k}{(2n+1)a\epsilon_{\rm M} }(n\Psi_{{\rm M}(n-1)}- (n+1)\Psi_{{\rm M}(n+1)})~,~~(n \ge 3).
\end{eqnarray*}
Hereafter, we set $\Psi_{{\rm M}(n)}=0$ for $n\geq 2$, because we assume
that the mother particles are CDM. For daughter particles, we obtain 
\begin{eqnarray*} 
\dot{\Psi}_{{\rm D}j0}&=&-\dfrac{q_{\rm D}k}{a\epsilon_{{\rm D}j} }\Psi_{{\rm D}j1}+\dfrac{1}{6}\dot{h}\dfrac{d\ln{f_{\rm D}^{(0)}}}{d\ln{q_{\rm D}}}+\dfrac{4\Gamma }{B_{\rm M}}\dfrac{q_{\rm D}^{\prime }}{q_{\rm D}}\dfrac{f_{\rm M}^{(0)}(q_{\rm D}^{\prime } )}{f_{{\rm D}}^{(0)}(q_{\rm D})}(\Psi_{\rm M0}(q_{\rm D}^{\prime })-\Psi _{{\rm D}j0}(q_{\rm D}))~,\\
\dot{\Psi}_{{\rm D}j1}&=&\dfrac{q_{\rm D}k}{3a\epsilon_{{\rm D}j} }(\Psi_{{\rm D}j0}- 2\Psi_{{\rm D}j2})+ \dfrac{4\Gamma }{B_{\rm M}}\dfrac{q_{\rm D}^{\prime }}{q_{\rm D}}\dfrac{f_{\rm M}^{(0)}(q_{\rm D}^{\prime } )}{f_{{\rm D}}^{(0)}(q_{\rm D})}(\Psi_{\rm M1}(q_{\rm D}^{\prime })-\Psi _{{\rm D}j1}(q_{\rm D}))~,\\
\dot{\Psi}_{{\rm D}j2}&=&\dfrac{q_{\rm D}k}{5a\epsilon_{{\rm D}j} }(2\Psi_{{\rm D}j1}- 3\Psi_{{\rm D}j3})-\bigg( \dfrac{1}{15}\dot{h}+\dfrac{2}{5}\dot{\eta }\bigg)\dfrac{d\ln{f_{{\rm D}}^{(0)}}}{d\ln {q_{\rm D}}}- \dfrac{4\Gamma }{B_{\rm M}}\dfrac{q_{\rm D}^{\prime }}{q_{\rm D}}\dfrac{f_{\rm M}^{(0)}(q_{\rm D}^{\prime } )}{f_{{\rm D}}^{(0)}(q_{\rm D})}\Psi _{{\rm D}j2}(q_{\rm D})~,\\
\dot{\Psi}_{{\rm D}j(n)}&=&\dfrac{q_{\rm D}k}{(2n+1)a\epsilon_{{\rm D}j} }(n\Psi_{{\rm D}j(n-1)}- (n+1)\Psi_{{\rm D}j(n+1)})- \dfrac{4\Gamma }{B_{\rm M}}\dfrac{q_{\rm D}^{\prime }}{q_{\rm D}}\dfrac{f_{\rm M}^{(0)}(q_{\rm D}^{\prime } )}{f_{{\rm D}}^{(0)}(q_{\rm D})}\Psi _{{\rm D}j(n)}(q_{\rm D})~~(n \ge 3).
\end{eqnarray*}
{In the limit $\Gamma\to 0$, the above equations reduce
to those of massive neutrinos in the standard perturbation theory
\cite{1995ApJ...455....7M}. Again, the terms which are related with the
decay process comes both from the collision term and the change of the
background evolution. For the moment equations of daugher particles
with $n\geq 2$, the decay terms only come from the change of the background
evolution, because the mother CDM particles do not have the corresponding
moments, i.e.  $\Psi_{{\rm M}(n)}=0$ for $n\geq 2$.}

\bibstyle{JHEP-2}
\bibliography{reference2011a}

\providecommand{\href}[2]{#2}\begingroup\raggedright\begin{thebibliography}{10}

\bibitem{Cold..Dark..Matter}
A.~Brett, {\it Cold dark matter},
\newblock Dundurn Press Ltd, 2005.

\bibitem{1999ApJ...524L..19M}
B.~{Moore}, S.~{Ghigna}, F.~{Governato}, G.~{Lake}, T.~{Quinn}, J.~{Stadel} and
  P.~{Tozzi}, {\it {Dark Matter Substructure within Galactic Halos}},  {\em
  \apjl} {\bf 524} (Oct., 1999) L19--L22
  [\href{http://arXiv.org/abs/arXiv:astro-ph/9907411}{{\tt
  arXiv:astro-ph/9907411}}].

\bibitem{2009ApJ...696.2115I}
T.~{Ishiyama}, T.~{Fukushige} and J.~{Makino}, {\it {Variation of the Subhalo
  Abundance in Dark Matter Halos}},  {\em \apj} {\bf 696} (May, 2009)
  2115--2125 [\href{http://arXiv.org/abs/0812.0683}{{\tt 0812.0683}}].

\bibitem{2003Sci...300.1909O}
J.~P. {Ostriker} and P.~{Steinhardt}, {\it {New Light on Dark Matter}},  {\em
  Science} {\bf 300} (June, 2003) 1909--1914
  [\href{http://arXiv.org/abs/arXiv:astro-ph/0306402}{{\tt
  arXiv:astro-ph/0306402}}].

\bibitem{2001ApJ...546L..77C}
R.~{Cen}, {\it {Decaying Cold Dark Matter Model and Small-Scale Power}},  {\em
  \apjl} {\bf 546} (Jan., 2001) L77--L80
  [\href{http://arXiv.org/abs/arXiv:astro-ph/0005206}{{\tt
  arXiv:astro-ph/0005206}}].

\bibitem{2004PhRvL..93g1302I}
K.~{Ichiki}, M.~{Oguri} and K.~{Takahashi}, {\it {Constraints from the
  Wilkinson Microwave Anisotropy Probe on Decaying Cold Dark Matter}},  {\em
  Physical Review Letters} {\bf 93} (Aug., 2004) 071302--+
  [\href{http://arXiv.org/abs/arXiv:astro-ph/0403164}{{\tt
  arXiv:astro-ph/0403164}}].

\bibitem{2010PhRvD..81h3511P}
A.~H.~G. {Peter}, {\it {Mapping the allowed parameter space for decaying dark
  matter models}},  {\em \prd} {\bf 81} (Apr., 2010) 083511--+
  [\href{http://arXiv.org/abs/1001.3870}{{\tt 1001.3870}}].

\bibitem{2003ApJ...597..645O}
M.~{Oguri}, K.~{Takahashi}, H.~{Ohno} and K.~{Kotake}, {\it {Decaying Cold Dark
  Matter and the Evolution of the Cluster Abundance}},  {\em \apj} {\bf 597}
  (Nov., 2003) 645--649
  [\href{http://arXiv.org/abs/arXiv:astro-ph/0306020}{{\tt
  arXiv:astro-ph/0306020}}].

\bibitem{2010PhRvD..81j3501P}
A.~H.~G. {Peter}, C.~E. {Moody} and M.~{Kamionkowski}, {\it {Dark-matter decays
  and self-gravitating halos}},  {\em \prd} {\bf 81} (May, 2010) 103501--+
  [\href{http://arXiv.org/abs/1003.0419}{{\tt 1003.0419}}].

\bibitem{2005PhRvD..72f3510K}
M.~{Kaplinghat}, {\it {Dark matter from early decays}},  {\em \prd} {\bf 72}
  (Sept., 2005) 063510--+
  [\href{http://arXiv.org/abs/arXiv:astro-ph/0507300}{{\tt
  arXiv:astro-ph/0507300}}].

\bibitem{1999PhRvD..60l3508K}
M.~{Kaplinghat}, R.~E. {Lopez}, S.~{Dodelson} and R.~J. {Scherrer}, {\it
  {Improved treatment of cosmic microwave background fluctuations induced by a
  late-decaying massive neutrino}},  {\em \prd} {\bf 60} (Dec., 1999) 123508--+
  [\href{http://arXiv.org/abs/arXiv:astro-ph/9907388}{{\tt
  arXiv:astro-ph/9907388}}].

\bibitem{1993NuPhB.403..671K}
M.~{Kawasaki}, G.~{Steigman} and H.-S. {Kang}, {\it {Cosmological evolution of
  an early-decaying particle}},  {\em Nuclear Physics B} {\bf 403} (Aug., 1993)
  671--706.

\bibitem{2009JCAP...06..005D}
S.~{DeLope Amigo}, W.~{Man-Yin Cheung}, Z.~{Huang} and S.-P. {Ng}, {\it
  {Cosmological constraints on decaying dark matter}},  {\em \jcap} {\bf 6}
  (June, 2009) 5--+ [\href{http://arXiv.org/abs/0812.4016}{{\tt 0812.4016}}].

\bibitem{2008PhRvD..77b3009A}
L.~A. {Anchordoqui}, A.~{Delgado}, C.~A. {Garc{\'{\i}}a Canal} and S.~J.
  {Sciutto}, {\it {Hunting long-lived gluinos at the Pierre Auger
  Observatory}},  {\em \prd} {\bf 77} (Jan., 2008) 023009--+
  [\href{http://arXiv.org/abs/0710.0525}{{\tt 0710.0525}}].

\bibitem{2005PhR...405..279B}
G.~{Bertone}, D.~{Hooper} and J.~{Silk}, {\it {Particle dark matter: evidence,
  candidates and constraints}},  {\em \physrep} {\bf 405} (Jan., 2005) 279--390
  [\href{http://arXiv.org/abs/arXiv:hep-ph/0404175}{{\tt
  arXiv:hep-ph/0404175}}].

\bibitem{2005PhLB..625....7K}
M.~{Kawasaki}, K.~{Kohri} and T.~{Moroi}, {\it {Hadronic decay of late-decaying
  particles and big-bang nucleosynthesis}},  {\em Physics Letters B} {\bf 625}
  (Oct., 2005) 7--12 [\href{http://arXiv.org/abs/arXiv:astro-ph/0402490}{{\tt
  arXiv:astro-ph/0402490}}].

\bibitem{2008PhLB..665...50P}
S.~{Palomares-Ruiz}, {\it {Model-independent bound on the dark matter
  lifetime}},  {\em Physics Letters B} {\bf 665} (July, 2008) 50--53
  [\href{http://arXiv.org/abs/0712.1937}{{\tt 0712.1937}}].

\bibitem{2011PhRvD..83f3504B}
N.~F. {Bell}, A.~J. {Galea} and R.~R. {Volkas}, {\it {Model for late dark
  matter decay}},  {\em \prd} {\bf 83} (Mar., 2011) 063504--+
  [\href{http://arXiv.org/abs/1012.0067}{{\tt 1012.0067}}].

\bibitem{1998PhRvL..81.4048C}
D.~J.~H. {Chung}, E.~W. {Kolb} and A.~{Riotto}, {\it {Nonthermal Supermassive
  Dark Matter}},  {\em Physical Review Letters} {\bf 81} (Nov., 1998)
  4048--4051 [\href{http://arXiv.org/abs/arXiv:hep-ph/9805473}{{\tt
  arXiv:hep-ph/9805473}}].

\bibitem{1999PhRvD..59d7301B}
K.~{Benakli}, J.~{Ellis} and D.~V. {Nanopoulos}, {\it {Natural candidates for
  superheavy dark matter in string and M theory}},  {\em \prd} {\bf 59} (Feb.,
  1999) 047301--+ [\href{http://arXiv.org/abs/arXiv:hep-ph/9803333}{{\tt
  arXiv:hep-ph/9803333}}].

\bibitem{2001PhRvL..86..954L}
W.~B. {Lin}, D.~H. {Huang}, X.~{Zhang} and R.~{Brandenberger}, {\it {Nonthermal
  Production of Weakly Interacting Massive Particles and the Subgalactic
  Structure of the Universe}},  {\em Physical Review Letters} {\bf 86} (Feb.,
  2001) 954--957 [\href{http://arXiv.org/abs/arXiv:astro-ph/0009003}{{\tt
  arXiv:astro-ph/0009003}}].

\bibitem{2002PhLB..527...18K}
H.~B. {Kim} and J.~E. {Kim}, {\it {Late decaying axino as CDM and its lifetime
  bound}},  {\em Physics Letters B} {\bf 527} (Feb., 2002) 18--22
  [\href{http://arXiv.org/abs/arXiv:hep-ph/0108101}{{\tt
  arXiv:hep-ph/0108101}}].

\bibitem{Nakamura:2010zzi}
{\bf Particle Data Group} Collaboration, K.~Nakamura {\em et.~al.}, {\it
  {Review of particle physics}},  {\em J. Phys.} {\bf G37} (2010) 075021.

\bibitem{2003PhRvD..67f3519B}
A.~{Bottino}, N.~{Fornengo} and S.~{Scopel}, {\it {Light relic neutralinos}},
  {\em \prd} {\bf 67} (Mar., 2003) 063519--+
  [\href{http://arXiv.org/abs/arXiv:hep-ph/0212379}{{\tt
  arXiv:hep-ph/0212379}}].

\bibitem{2003PhRvD..68d3506B}
A.~{Bottino}, F.~{Donato}, N.~{Fornengo} and S.~{Scopel}, {\it {Lower bound on
  the neutralino mass from new data on CMB and implications for relic
  neutralinos}},  {\em \prd} {\bf 68} (Aug., 2003) 043506--+
  [\href{http://arXiv.org/abs/arXiv:hep-ph/0304080}{{\tt
  arXiv:hep-ph/0304080}}].

\bibitem{2010arXiv1009.0549B}
A.~V. {Belikov}, J.~F. {Gunion}, D.~{Hooper} and T.~M.~P. {Tait}, {\it {CoGeNT,
  DAMA, and Light Neutralino Dark Matter}},  {\em ArXiv e-prints} (Sept., 2010)
  [\href{http://arXiv.org/abs/1009.0549}{{\tt 1009.0549}}].

\bibitem{2003AJ....125...32B}
M.~{Bernardi}, R.~K. {Sheth}, M.~{SubbaRao}, G.~T. {Richards}, S.~{Burles},
  A.~J. {Connolly}, J.~{Frieman}, R.~{Nichol}, J.~{Schaye}, D.~P. {Schneider},
  D.~E. {Vanden Berk}, D.~G. {York}, J.~{Brinkmann} and D.~Q. {Lamb}, {\it {A
  Feature at z\~{}3.2 in the Evolution of the Ly{$\alpha$} Forest Optical
  Depth}},  {\em \aj} {\bf 125} (Jan., 2003) 32--52
  [\href{http://arXiv.org/abs/arXiv:astro-ph/0206293}{{\tt
  arXiv:astro-ph/0206293}}].

\bibitem{2009JCAP...05..012B}
A.~{Boyarsky}, J.~{Lesgourgues}, O.~{Ruchayskiy} and M.~{Viel}, {\it
  {Lyman-{$\alpha$} constraints on warm and on warm-plus-cold dark matter
  models}},  {\em \jcap} {\bf 5} (May, 2009) 12--+
  [\href{http://arXiv.org/abs/0812.0010}{{\tt 0812.0010}}].

\bibitem{2000astro.ph..2400Z}
H.~{Ziaeepour}, {\it {Cosmic Equation of State, Quintessence and Decaying Dark
  Matter}},  {\em ArXiv Astrophysics e-prints} (Feb., 2000)
  [\href{http://arXiv.org/abs/arXiv:astro-ph/0002400}{{\tt
  arXiv:astro-ph/0002400}}].

\bibitem{2004PhRvD..69f3512Z}
H.~{Ziaeepour}, {\it {Quintessence from the decay of superheavy dark matter}},
  {\em \prd} {\bf 69} (Mar., 2004) 063512--+
  [\href{http://arXiv.org/abs/arXiv:astro-ph/0308515}{{\tt
  arXiv:astro-ph/0308515}}].

\bibitem{2011ApJS..192...14J}
N.~{Jarosik}, C.~L. {Bennett}, J.~{Dunkley}, B.~{Gold}, M.~R. {Greason},
  M.~{Halpern}, R.~S. {Hill}, G.~{Hinshaw}, A.~{Kogut}, E.~{Komatsu},
  D.~{Larson}, M.~{Limon}, S.~S. {Meyer}, M.~R. {Nolta}, N.~{Odegard},
  L.~{Page}, K.~M. {Smith}, D.~N. {Spergel}, G.~S. {Tucker}, J.~L. {Weiland},
  E.~{Wollack} and E.~L. {Wright}, {\it {Seven-year Wilkinson Microwave
  Anisotropy Probe (WMAP) Observations: Sky Maps, Systematic Errors, and Basic
  Results}},  {\em \apjs} {\bf 192} (Feb., 2011) 14--+
  [\href{http://arXiv.org/abs/1001.4744}{{\tt 1001.4744}}].

\bibitem{2010MNRAS.401.2148P}
W.~J. {Percival}, B.~A. {Reid}, D.~J. {Eisenstein}, N.~A. {Bahcall},
  T.~{Budavari}, J.~A. {Frieman}, M.~{Fukugita}, J.~E. {Gunn}, {\v
  Z}.~{Ivezi{\'c}}, G.~R. {Knapp}, R.~G. {Kron}, J.~{Loveday}, R.~H. {Lupton},
  T.~A. {McKay}, A.~{Meiksin}, R.~C. {Nichol}, A.~C. {Pope}, D.~J. {Schlegel},
  D.~P. {Schneider}, D.~N. {Spergel}, C.~{Stoughton}, M.~A. {Strauss}, A.~S.
  {Szalay}, M.~{Tegmark}, M.~S. {Vogeley}, D.~H. {Weinberg}, D.~G. {York} and
  I.~{Zehavi}, {\it {Baryon acoustic oscillations in the Sloan Digital Sky
  Survey Data Release 7 galaxy sample}},  {\em \mnras} {\bf 401} (Feb., 2010)
  2148--2168 [\href{http://arXiv.org/abs/0907.1660}{{\tt 0907.1660}}].

\bibitem{2005ApJ...633..560E}
D.~J. {Eisenstein}, I.~{Zehavi}, D.~W. {Hogg}, R.~{Scoccimarro}, M.~R.
  {Blanton}, R.~C. {Nichol}, R.~{Scranton}, H.-J. {Seo}, M.~{Tegmark},
  Z.~{Zheng}, S.~F. {Anderson}, J.~{Annis}, N.~{Bahcall}, J.~{Brinkmann},
  S.~{Burles}, F.~J. {Castander}, A.~{Connolly}, I.~{Csabai}, M.~{Doi},
  M.~{Fukugita}, J.~A. {Frieman}, K.~{Glazebrook}, J.~E. {Gunn}, J.~S.
  {Hendry}, G.~{Hennessy}, Z.~{Ivezi{\'c}}, S.~{Kent}, G.~R. {Knapp}, H.~{Lin},
  Y.-S. {Loh}, R.~H. {Lupton}, B.~{Margon}, T.~A. {McKay}, A.~{Meiksin}, J.~A.
  {Munn}, A.~{Pope}, M.~W. {Richmond}, D.~{Schlegel}, D.~P. {Schneider},
  K.~{Shimasaku}, C.~{Stoughton}, M.~A. {Strauss}, M.~{SubbaRao}, A.~S.
  {Szalay}, I.~{Szapudi}, D.~L. {Tucker}, B.~{Yanny} and D.~G. {York}, {\it
  {Detection of the Baryon Acoustic Peak in the Large-Scale Correlation
  Function of SDSS Luminous Red Galaxies}},  {\em \apj} {\bf 633} (Nov., 2005)
  560--574 [\href{http://arXiv.org/abs/arXiv:astro-ph/0501171}{{\tt
  arXiv:astro-ph/0501171}}].

\bibitem{2007MNRAS.381.1053P}
W.~J. {Percival}, S.~{Cole}, D.~J. {Eisenstein}, R.~C. {Nichol}, J.~A.
  {Peacock}, A.~C. {Pope} and A.~S. {Szalay}, {\it {Measuring the Baryon
  Acoustic Oscillation scale using the Sloan Digital Sky Survey and 2dF Galaxy
  Redshift Survey}},  {\em \mnras} {\bf 381} (Nov., 2007) 1053--1066
  [\href{http://arXiv.org/abs/0705.3323}{{\tt 0705.3323}}].

\bibitem{2011ApJ...730..119R}
A.~G. {Riess}, L.~{Macri}, S.~{Casertano}, H.~{Lampeitl}, H.~C. {Ferguson},
  A.~V. {Filippenko}, S.~W. {Jha}, W.~{Li} and R.~{Chornock}, {\it {A 3\%
  Solution: Determination of the Hubble Constant with the Hubble Space
  Telescope and Wide Field Camera 3}},  {\em \apj} {\bf 730} (Apr., 2011)
  119--+ [\href{http://arXiv.org/abs/1103.2976}{{\tt 1103.2976}}].

\bibitem{1996ApJ...473..576F}
D.~J. {Fixsen}, E.~S. {Cheng}, J.~M. {Gales}, J.~C. {Mather}, R.~A. {Shafer}
  and E.~L. {Wright}, {\it {The Cosmic Microwave Background Spectrum from the
  Full COBE FIRAS Data Set}},  {\em \apj} {\bf 473} (Dec., 1996) 576--+
  [\href{http://arXiv.org/abs/arXiv:astro-ph/9605054}{{\tt
  arXiv:astro-ph/9605054}}].

\bibitem{2003ApJ...591L.107S}
F.~J. {S{\'a}nchez-Salcedo}, {\it {Unstable Cold Dark Matter and the Cuspy Halo
  Problem in Dwarf Galaxies}},  {\em \apjl} {\bf 591} (July, 2003) L107--L110
  [\href{http://arXiv.org/abs/arXiv:astro-ph/0305496}{{\tt
  arXiv:astro-ph/0305496}}].

\bibitem{1990eaun.book.....K}
E.~W. {Kolb} and M.~S. {Turner}, {\em {The early universe.}}
\newblock 1990.

\bibitem{2011ApJS..192...18K}
E.~{Komatsu}, K.~M. {Smith}, J.~{Dunkley}, C.~L. {Bennett}, B.~{Gold},
  G.~{Hinshaw}, N.~{Jarosik}, D.~{Larson}, M.~R. {Nolta}, L.~{Page}, D.~N.
  {Spergel}, M.~{Halpern}, R.~S. {Hill}, A.~{Kogut}, M.~{Limon}, S.~S. {Meyer},
  N.~{Odegard}, G.~S. {Tucker}, J.~L. {Weiland}, E.~{Wollack} and E.~L.
  {Wright}, {\it {Seven-year Wilkinson Microwave Anisotropy Probe (WMAP)
  Observations: Cosmological Interpretation}},  {\em \apjs} {\bf 192} (Feb.,
  2011) 18--+ [\href{http://arXiv.org/abs/1001.4538}{{\tt 1001.4538}}].

\bibitem{1995ApJ...455....7M}
C.-P. {Ma} and E.~{Bertschinger}, {\it {Cosmological Perturbation Theory in the
  Synchronous and Conformal Newtonian Gauges}},  {\em \apj} {\bf 455} (Dec.,
  1995) 7--+ [\href{http://arXiv.org/abs/arXiv:astro-ph/9506072}{{\tt
  arXiv:astro-ph/9506072}}].

\end{thebibliography}\endgroup

\end{document}